\documentclass[
10pt,
aps,
pra,
twocolumn,
groupedaddress,
superscriptaddress,
amsmath,
amssymb,
reprint,
longbibliography,
floatfix]{revtex4-2}

\usepackage{natbib}
\usepackage{array}
\usepackage[hidelinks]{hyperref}

\usepackage{tabularx}

\usepackage{mathtools}
\usepackage{cancel}
\usepackage{braket}

\usepackage{enumitem}
\usepackage[utf8x]{inputenc}
\usepackage{graphicx}
\usepackage{dcolumn}
\usepackage{bm}

\usepackage[switch]{lineno}
\usepackage{subfigure}
\usepackage{tikz}
\usepackage{multirow}
\usepackage{placeins}

\usetikzlibrary{arrows.meta}
\usetikzlibrary{arrows,decorations.pathmorphing,backgrounds,positioning,fit,petri}

\begin{document}

\title{A first-principles linear response theory for open quantum systems and its application to Orbach and direct magnetic relaxation in Ln-based coordination polymers}

\author{Mikołaj Żychowicz}
\email{mikolaj.zychowicz@uj.edu.pl}
\author{Jakub J. Zakrzewski}
\author{Szymon Chorazy}
\email{simon.chorazy@uj.edu.pl}
\affiliation{Faculty of Chemistry, Jagiellonian University, Gronostajowa 2, 30-387 Krakow, Poland}
\author{Alessandro Lunghi}
\email{lunghia@tcd.ie}
\affiliation{School of Physics, AMBER and CRANN Institute, Trinity College, Dublin 2, Ireland}

\begin{abstract}
Single-Molecule Magnets (SMMs) exhibit slow magnetic relaxation as a result of axial magnetic anisotropy inhibiting spin-phonon transitions. In order to establish a direct link between physical observables and the microscopic theory of magnetic relaxation, we here develop and numerically implement a first-principles linear-response theory for open quantum systems that provides access to the complex a.c. magnetic susceptibility in the presence of an oscillating a.c. magnetic field. Once combined with density functional theory and multiconfigurational electronic structure simulations, this formalism is applied in a fully first-principles fashion to three cyanido-bridged Ln/Y-based coordination polymers with general formula \( \{ \mathrm{Ln}_{x}^{\text{III}}\mathrm{Y}_{1-x}^{\text{III}}[ \mathrm{Co}^{\text{III}}(\mathrm{CN})_6 ]\}\), where Ln $=$ Yb (\textbf{1}), Tb (\textbf{2}), and Dy (\textbf{3}). The method is able to reproduce the low-temperature direct relaxation process and its field dependence, as well as the high-temperature Orbach relaxation regime for all the investigated compounds. These results demonstrate the feasibility of ab initio simulations of magnetic a.c. susceptibility in lanthanide-based SMMs and support the potential of further development of ab initio open quantum systems methods towards the completion of a magnetization dynamics theory.
\end{abstract}

\maketitle

\twocolumngrid

\section*{Introduction}

Single-Molecule Magnets (SMMs) occupy a central position among various types of nanomagnets and have attracted a great deal of scientific attention over the years because they exhibit a slow magnetic relaxation of purely molecular origin. \cite{Gatteschi2006,Woodruff2013,Chilton2022} Their ability to retain magnetization below a certain blocking temperature has made them attractive in the context of ultra-high-density data storage, \cite{Mannini2009} and spintronics applications.\cite{Candini2011,Coronado2020} In recent years, lanthanide-based SMMs have pushed these prospects considerably further, with effective barriers of magnetization reversal reaching 1843 cm$^{-1}$ and soft magnetic hysteresis persisting up to 100 K. \cite{Emerson2025}

Because the performance of a SMM is determined by how slowly its magnetization relaxes, the central experimental quantity is the characteristic magnetic relaxation time $\tau$. Since the earliest experiments and their interpretation, \cite{DeHaas1938,Casimir1938} alternating-current (a.c.) measurements of the complex magnetic susceptibility have gradually become the standard method for probing this dynamics in molecular nanomagnets \cite{Sessoli1993} such as SMMs, single-chain magnets, and other low-dimensional magnetic systems, as well as a wide variety of correlated systems. \cite{Coulon2006,Balanda2013,Topping2019,ZabalaLekuona2021} Currently, in commercial SQUID magnetometers \cite{QD_ACsus_and_specs} and inductive susceptometers \cite{Michelena2017, Riordan2019}, a small oscillating field of amplitude typically a few Oe and frequency in the $10^{1}$-$10^{4}$~Hz range (PPMS/ACMS II system) is superimposed on a static bias (d.c.) field. The resulting time-dependent magnetic moment is detected by pickup coils and a SQUID sensor with extremely high sensitivity. Compared to direct d.c. relaxation measurements, a.c. susceptibility is both technically simpler and far more sensitive to small changes in magnetization. It also remains applicable even when the relaxation is too fast to be resolved in a conventional magnetization $M(t)$ decay experiment, e.g., when the relaxation time is short enough to interfere with the time scale needed to remove the magnetic field in the experiment. \cite{Topping2019,Balanda2013} In practice, most relaxation times $\tau$ reported for SMMs are extracted by fitting the measured $\chi'(\omega)$ and $\chi''(\omega)$ curves with Debye-type phenomenological models and their generalizations. \cite{Debye1929,ColeCole1941,Davidson1951,HavriliakNegami1967}

The central role of $\tau$ in experiments has naturally motivated substantial effort towards its ab initio simulation. State-of-the-art theoretical approaches are typically based on some flavor of time-dependent perturbation theory applied to an open quantum system composed of the magnetic spin degrees of freedom weakly coupled to a Markovian phonon bath. \cite{Lunghi2017_1,Lunghi2017_2,Goodwin2017,Ho2018,Lunghi2019,Briganti2021,Mondal2022,Lunghi2022,Kragskow2023,Lunghi2023} These methods have been highly successful in rationalizing relaxation pathways and trends across many families of molecular magnets. At the same time, they usually treat the problem as a field-free master-equation dynamics for the reduced density matrix and infer the relaxation time indirectly, most often from the smallest non-zero eigenvalues of an effective rate matrix or by fitting the exponential decay of magnetization of a previously axially magnetized state, typically within a secular approximation. \cite{Lunghi2023} In this sense, they are formally much closer to the determination of longitudinal $T_1$ relaxation times in electron paramagnetic resonance (EPR) or to d.c. magnetization remanence experiments, where one follows free relaxation after a static perturbation has been removed. \cite{AbragamBleaney1970} In other words, they completely disregard the applied oscillating magnetic field and its effect during the a.c. experiments, since the dynamics is characterized by free decay in the absence of a driving field.

This reveals a clear conceptual gap. Although a.c. susceptibility is the experimentally dominant source of relaxation times in the SMM field; essentially, all current microscopic methodologies still simulate rates first and compare them only indirectly with $\tau$ values extracted from fitted a.c. data. The oscillating magnetic field itself, together with the fact that the experimentally measured object is the full complex susceptibility $\chi(\omega)$ rather than an isolated rate constant, is therefore not treated as the central quantity of the theory. As a consequence, information about how a specific perturbing operator and measured observable select and weight different relaxation channels is lost already at the theoretical level. Closing this gap is important not only for a more direct comparison with experiment but also for putting ab initio theory and experimental a.c. analysis on the same footing and for enabling genuinely predictive, fully in silico studies of slow magnetic relaxation.

\begin{figure}[htbp!]
    \centering
    \includegraphics[scale=1]{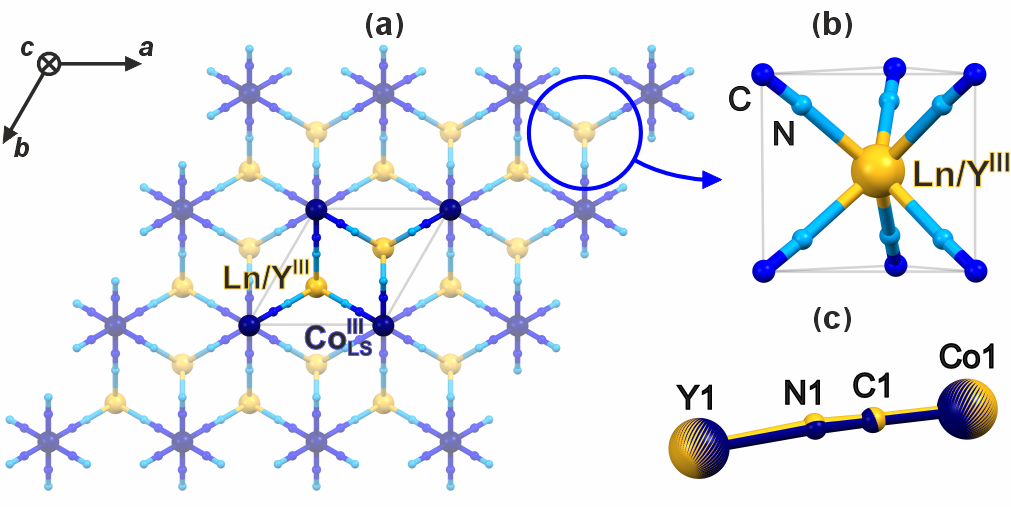}
    \caption{Experimental crystal structure of the anhydrous Ln$^{\mathrm{III}}$/Y$^{\mathrm{III}}$-[$\mathrm{Co}^{\mathrm{III}}(\mathrm{CN})_6$]$^{3-}$ three-dimensional framework: (a) view of the $3\times3\times2$ supercell along the crystallographic $c$ axis, with the unit cell highlighted; (b) geometry of the Ln/Y center surrounded by six cyanido ligands; (c) overlap between the experimental and optimized asymmetric units for the Y-based framework, shown in deep blue and yellow, respectively. In panel (a), the $\mathrm{Co}^{\mathrm{III}}_{\mathrm{LS}}$ label emphasizes the diamagnetic character of the low-spin cobalt(III) centers.}
    \label{fig:1}
    \hfill
\end{figure}

In this contribution, we address this gap by presenting a complete first principles derivation and fully ab initio implementation of a theoretical methodology based on causal, retarded linear response theory for open quantum systems \cite{Uchiyama2009} in the presence of an oscillating magnetic field for complex susceptibility simulation, going even beyond the Born-Markov \cite{BreuerPetruccione2002} level of approximations. Instead of inferring relaxation times from field-free dissipative dynamics, our framework gives direct access to the complex a.c. magnetic susceptibility itself and thereby establishes a quantitative bridge between microscopic spin-phonon coupling and the experimentally recorded $\chi'(\omega)$ and $\chi''(\omega)$, allowing an equal treatment of both simulation and experiment using phenomenological models. We demonstrate the accuracy of this methodology by reporting numerical results for a family of lanthanide-based SMMs embedded in the three-dimensional cyanido-bridged coordination polymer matrix diluted with diamagnetic Y(III) metal centers with general formula \( \{ \mathrm{Ln}_{x}^{\text{III}}\mathrm{Y}_{1-x}^{\text{III}}[ \mathrm{Co}^{\text{III}}(\mathrm{CN})_6 ]\}\), where Ln $=$ Yb (\textbf{1}), Tb (\textbf{2}), Dy (\textbf{3}) and x $=$ 0.05 (\textbf{1}), 0.025 (\textbf{2}), 0.12 (\textbf{3}) respectively (see Figure \ref{fig:1}). \cite{Xin2019, Wang2023}

We structure the article so that, for clarity, the main text contains only the central theoretical results and physical conclusions, while we provide all the necessary step-by-step derivations in a self-contained, comprehensive Supplementary Information (SI) for the interested reader to follow. In the Theory section, we first introduce a.c. susceptibility and its phenomenological modeling, then revisit linear response for closed quantum systems and generalize it to open quantum systems, and finally specialize the formalism to linear vibronic spin-phonon coupling with a harmonic phonon bath. Then, we describe the numerical implementation and computational protocol, including the ab initio evaluation of spin-phonon coupling coefficients. Next, we provide a brief section on magnetic relaxation time, introducing the previous computational methodology \cite{Lunghi2022} for its simulation together with the recipe for its extraction from experimental or ab initio a.c. susceptibility data and the phenomenological model used for the mechanistic interpretation. We close the article with a presentation of the results for the selected coordination polymers, including a comparison with existing methodologies for computing magnetic relaxation time and a critical discussion of their differences and remaining challenges.

\section*{Theory}

{\bf A.C. susceptibility.}
In the small-field, linear response regime relevant to the measurements under an oscillating field in SQUID magnetometers, we can define frequency-dependent magnetic a.c. susceptibility $\chi_\mathrm{a.c.}(\omega)$ as
\begin{equation}\label{eq:ac-sus-def}
     \chi_\mathrm{a.c.}(\omega)=\frac{M_\mathrm{a.c.}(\omega)}{H_\mathrm{a.c.}(\omega)},
\end{equation}
where magnetization $M_\mathrm{a.c.}(\omega) =  \tilde{M}_\mathrm{a.c.}(\omega)e^{-i(\omega t -\theta(\omega))}$ and the applied magnetic field $ H_\mathrm{a.c.}(\omega) =   H_{a.c.}e^{-i\omega t}$ are treated as frequency-dependent complex quantities to accommodate the possible phase shift $ \theta(\omega)$ (see Section S2). \cite{Boca1999,Balanda2013,Topping2019} We note that the shift, together with the amplitude $\tilde{M}_\mathrm{a.c.}(\omega)$, is also a function of the field frequency and represents a slow relaxation process that takes place in the sample. At the same time, the real-time-dependent measured quantities are given as $M_\mathrm{a.c.}(t) = \mathrm{Re}\left(  M_\mathrm{a.c.}(\omega) \right), H_\mathrm{a.c.}(t) = \mathrm{Re}\left(H_\mathrm{a.c.}(\omega) \right).$ Since, in general, the magnetic susceptibility is an anisotropic second-rank tensor with regard to the direction of vectors of the magnetic field and the resultant magnetization, we omit that in the notation due to the common experimental setup, i.e., we are only interested in a fixed field direction such that $\vec{H}(t)$ and $\vec{M}(t)$ are collinear and, together with $\bar{\chi}_\mathrm{a.c.}(\omega)$, can be treated as scalars parametrized by a single orientation of a.c. field. Furthermore, in the linear regime, we approximate the differential definition of susceptibility (S2.2) discussed in the SI with a finite difference quotient \eqref{eq:ac-sus-def} as it is also experimentally accessed as the finite ratio. \cite{Kahn2021} We also note that since we are only interested here in magnetically diluted materials with $\chi\ll1$, the overall magnetic induction $B=\mathrm{\mu_0}(H+M)$ characterizing the magnetic field inside the substance, where $\mathrm{\mu_0}$ is the magnetic permeability of vacuum, is assumed to always be approximately $B\approx\mathrm{\mu_0}H$, thus neglecting the contribution from magnetization of the sample itself. Under these assumptions, a complex a.c. susceptibility can be conveniently represented as
\begin{equation}\label{eq:chi-complex}
    \chi_\mathrm{a.c.}(\omega) = \chi'(\omega) + i \chi''(\omega),
\end{equation}
where the quantities $\chi'(\omega)$ and $\chi''(\omega)$ are conventionally referred
to as the in-phase (dispersive) and out-of-phase (dissipative) components of
the susceptibility, \cite{Gorter1936} respectively, and we used the "+" sign convention here.

More generally, within linear-response theory, a generalized observable (displacement) $X(t)$
responding to a time-dependent generalized force $F_\mathrm{A}(t)$ can be written as
\begin{equation}\label{eq:lr-convolution}
    X(t) = \int_{-\infty}^{t} \tilde{\phi}_\mathrm{AX}(t-t')F_\mathrm{A}(t')dt',
\end{equation}
where $\tilde{\phi}_\mathrm{AX}(t-t')$ is a causal response function that should vanish for $t<t'$. We can remove this requirement by introducing the Heaviside step function $\Theta(t)$ and write $\tilde{\phi}_\mathrm{AX}(t)=\Theta(t)\phi_\mathrm{AX}(t)$ using the unrestricted response function $\phi_\mathrm{AX}(t)$. Then, as shown in Section S2 (S2.30 and S2.31), for time-translation-invariant systems, the Fourier transform (FT)
\begin{equation}\label{eq:sus-omega-heaviside}
    \chi_\mathrm{AX}(\omega) = \int_{-\infty}^{\infty}\Theta(t)\phi_{AX}(t)e^{i\omega t}dt=\int_{0}^{\infty}\phi_\mathrm{AX}(t)e^{i\omega t}dt
\end{equation}
defines a frequency-dependent susceptibility, and \eqref{eq:ac-sus-def} is recovered as a special case with $\chi_\mathrm{AX} \equiv \chi_\mathrm{a.c.}$, $F_\mathrm{A} \equiv H$ and $X \equiv M$, as \eqref{eq:lr-convolution} has the form of a convolution. \cite{Kubo1957,Kubo1966} Because $\chi_\mathrm{AX}(\omega)$ is the Fourier transform of a causal function, its real and imaginary parts are connected by the Kramers-Kronig relations, \cite{Kramers1928,Kronig1926,Nussenzveig1972} so $\chi'(\omega)$ and $\chi''(\omega)$ are not strictly independent quantities. To model them phenomenologically, one can define time-dependent magnetization
\begin{equation}\label{eq:mag-adiabatic}
    M(t) = \chi_\mathrm{S} H(t) + m(t)
\end{equation}
as a sum of an instantaneous (adiabatic) response $\chi_\mathrm{S} H(t)$ that follows the field immediately on the investigated time scale, and a slow-relaxing part $m(t)$. Then we can use the simplest description of slow relaxation using a first-order differential equation for $m(t)$
\begin{equation}\label{eq:mag-slow-diff}
    \tau \frac{dm(t)}{dt} + m(t) = (\chi_T - \chi_S) H(t),
\end{equation}
which yields an exponential decay of $m(t)$ with a single relaxation time $\tau$ after removing the field $H(t)=0$. For a harmonic drive $H(t)=H_\mathrm{ac}\cos(\omega t)$, \eqref{eq:mag-slow-diff} leads to the well-known Debye \cite{Debye1929} expression, see (S2.44), originally developed in the context of dielectric relaxation,
\begin{equation}\label{eq:debye-model}
    \chi_\mathrm{D}(\omega) = \chi_\mathrm{S} + \frac{\chi_\mathrm{T} - \chi_\mathrm{S}}{1 - i \omega \tau},
\end{equation}
which satisfies correct boundary conditions with $\chi(\omega) \to \chi_\mathrm{S}$ and $\chi(\omega) \to \chi_\mathrm{T}$ as $\omega \to \infty$ and $\omega \to 0$, respectively. Those represent the isothermal susceptibility $\chi_\mathrm{T}$ when the system has time to equilibrate with the environment, and the adiabatic $\chi_\mathrm{S}$ for the frequency regime where it cannot exchange energy with the bath on the timescale of the field oscillations. The Debye model assumes a unique relaxation time and purely exponential relaxation. However, real solid-state systems, and in particular single-molecule magnets, typically exhibit broad distributions of relaxation times arising from structural disorder, multiple magnetic centers, anisotropic environments, and additional (overlapping) relaxation channels. \cite{Ho2016} This is why various empirical generalizations of \eqref{eq:debye-model} have been developed \cite{ColeCole1941,Davidson1951} and used for experimental data analysis, of which one of the broadest and most universal is
\begin{equation}\label{eq:hn-model}
    \chi(\omega) = \chi_\mathrm{S} + \frac{\chi_\mathrm{T}-\chi_\mathrm{S}}{(1-(i\tau\omega)^{1-\alpha})^{\beta}}
\end{equation}
known as the Havriliak–Negami (HN) model. \cite{HavriliakNegami1967} Detailed derivations of the presented equations in connection to the complex susceptibility, linear response, extended with the Green's functions approach, and phenomenological models are given in Section S2 of the SI.

\begin{figure*}[htbp!]
    \centering
    \includegraphics[scale=0.91]{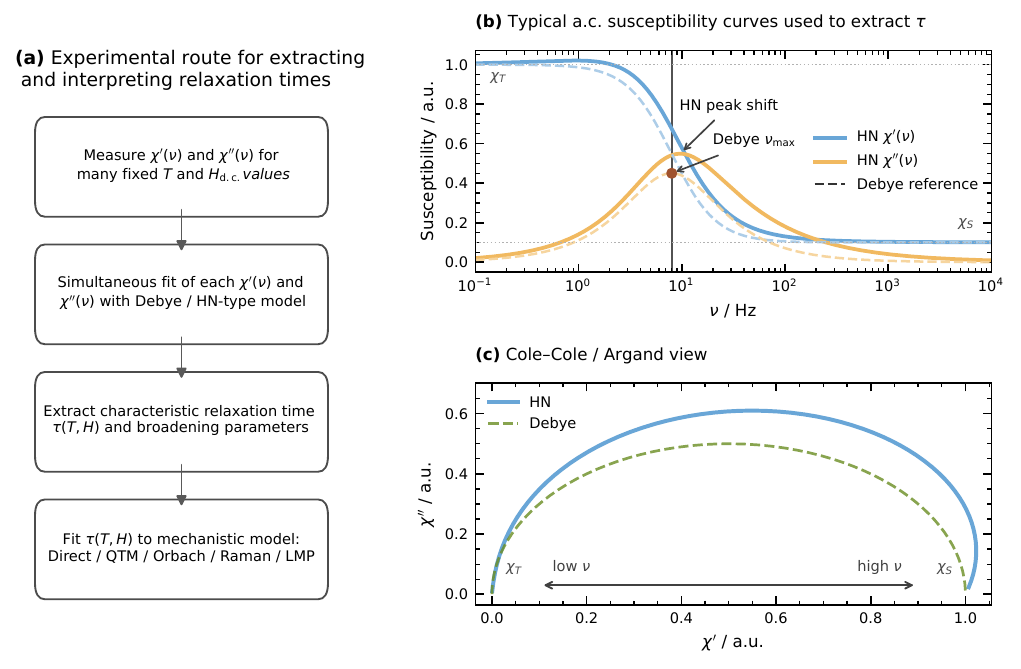}
    \caption{Scheme of usual experimental workflow for extracting magnetic relaxation times from experiment (a) with typical $\chi'(\nu)$ and $\chi''(\nu)$ traces of Debye model \eqref{eq:debye-model} compared to HN model \eqref{eq:hn-model} with $\alpha$ = 0.2, $\beta$ = 0.8 (both for the same $\tau$ = 0.02 s) and the corresponding Cole-Cole plot (c) showing the effect of broadening phenomenological HN parameters.}
    \label{fig:0}
    \hfill
\end{figure*}

In essentially all current analyses, the a.c. susceptibility is treated as a phenomenological input function of $\omega$; relaxation times and model parameters are extracted by fitting \eqref{eq:hn-model} to experimental $\chi'(\omega)$ and $\chi''(\omega)$. \cite{Boudalis2020,Liberka2022} A typical experimental workflow for extracting $\tau$ as a function of temperature and static d.c. magnetic field is presented in panel (a) of Figure \ref{fig:0}.
Although the characteristic magnetic relaxation time for the simple Debye model \eqref{eq:debye-model} can be intuitively read from the position of maxima on the $\chi''$ plot versus the linear frequency $\nu$ as $\tau = 1/(2\mathrm{\pi}\nu_{\mathrm{max}})$, traces encoding a distribution of relaxation times described by the general HN model \eqref{eq:hn-model} need to be precisely fit to extract this information as demonstrated by the peak shift induced by the phenomenological broadening parameters $\alpha$ and $\beta$ (see panels (b) and (c) of Figure \ref{fig:0}). The real $\chi'$ component, which has a deflection point at the position of $\chi''$ maxima, offers a straightforward access to the values of isothermal $\chi_\mathrm{T}$ and adiabatic $\chi_\mathrm{S}$ susceptibility values in the limits of low and high frequency, respectively, while their difference $\chi_\mathrm{T}-\chi_\mathrm{S}$ effectively scales the values of $\chi''$. The extracted relaxation times can then be interpreted with phenomenological mechanistic models as presented in the discussed scheme or compared to the results of the master-equation type of ab initio simulations for $\tau$. We cover the specifics of such approaches used in this work in the Computational Methods - Modeling of magnetic relaxation times subsection in connection with the employed software.

Taking into account the above, the central aim of this work is to compute the complex susceptibility $\chi(\omega)$ itself from first principles by evaluating the true linear response of an open quantum system to an oscillating magnetic field and only then to use a phenomenological model to extract relaxation times mimicking the experimental workflow. This provides a direct quantitative bridge between microscopic spin-phonon dynamics and the experimentally measured a.c. susceptibility. In this regard, the first step is to find a working expression for the response function $\chi_\mathrm{AX}(\omega)$ of an open quantum system connected to a thermal bath to further use \eqref{eq:sus-omega-heaviside} to find the susceptibility formula.

{\bf Linear response for a closed quantum system.}
For reference, we briefly summarize the standard quantum-mechanical expression for the response function, deferring all derivations to Section S3 in the SI. We consider a time-independent Hamiltonian $H_0$ perturbed by a weak, time-dependent term of the form
\begin{equation}\label{eq:h-tot}
    H_\mathrm{tot}(t) = H_0 + \delta H_\mathrm{pert}(t) = H_0 - AF_\mathrm{A}(t),
\end{equation}
where $A$ is a system (coupling) operator and $F_A(t)$ is a generalized force. \cite{Sakurai1994} The system is assumed to be in thermal equilibrium with respect to $H_0$ as $t\to -\infty$, with the density operator $\rho_\mathrm{eq} = e^{-\beta H_0}/Tr(e^{-\beta H_0})$, $\beta=1/(\mathrm{k_B}T)$ (adiabatic perturbation switching). For an observable $X$, the first-order change of its expectation value $\delta\braket{X(t)}$ induced by a small perturbation $\delta H_\mathrm{pert}(t)=- AF_\mathrm{A}(t)$ can be written in the convolution form
\begin{equation}\label{eq:lr-closed-convolution}
    \delta\braket{X(t)} = \int_{-\infty}^{t} \frac{i}{\mathrm{\hbar}}\braket{[X(t-t'),A]}_\mathrm{eq}F_\mathrm{A}(t')dt'
\end{equation}
where $\braket{O}_\mathrm{eq} = Tr(\rho_\mathrm{eq}O)$ and we used the stationary time-translation invariance condition with respect to $H_0$ to write the expectation value of the commutator as $\braket{[X(t-t'),A]}_\mathrm{eq}=\braket{[X(t),A(t')]}_\mathrm{eq}$ with $A\equiv A(0)$. Since \eqref{eq:lr-closed-convolution} is the quantum analog of \eqref{eq:lr-convolution} we can identify the corresponding unrestricted response function as an expectation value of the commutator given by the Kubo formula \cite{Kubo1957,Kubo1966}
\begin{equation}\label{eq:kubo-response}
    \phi_\mathrm{AX}(t) = \frac{i}{\mathrm{\hbar}}\braket{[X(t),A]}_\mathrm{eq}=\frac{i}{\mathrm{\hbar}}\mathrm{Tr}\bigl(Xe^{-i\mathcal{L}_\mathrm{eq}t}\{[A,\rho_\mathrm{eq}]\}\bigr),
\end{equation}
where $X(t)=e^{iH_0 t/\mathrm{\hbar}}Xe^{-iH_0 t/\mathrm{\hbar}}$ is the Heisenberg-picture operator with respect to $H_0$ and we gave it also in the Liouville super-operator form $\mathcal{L}_\mathrm{eq}O = \frac{1}{\mathrm{\hbar}}[H_0,O]$ as it will be used later. The corresponding frequency-dependent susceptibility is obtained as the one-sided Fourier transform of \eqref{eq:kubo-response}
\begin{equation}\label{eq:kubo-sus}
    \begin{aligned}
    \chi_\mathrm{AX}(\omega) &= \int_{0}^{\infty} \phi_\mathrm{AX}(t)e^{i\omega t}dt = \int_{-\infty}^{\infty} \Theta(t)\phi_\mathrm{AX}(t)e^{i\omega t}dt \\ &= \frac{i}{\mathrm{\hbar}}\int_{-\infty}^{\infty} \Theta(t) \braket{[X(t),A]}_\mathrm{eq}e^{i\omega t}dt,    
    \end{aligned}
\end{equation}
which is the close-system counterpart of \eqref{eq:sus-omega-heaviside} with the Heaviside step function that enforces causality. \cite{Kubo1991} In the magnetic case considered below, $A$ is identified with the magnetic dipole moment operator $\mu$ in a given direction and $F_A(t)$ with the applied magnetic field, so that $\chi_\mathrm{AX}(\omega)$ for an oscillating field reduces to the magnetic susceptibility $\chi_\mathrm{a.c.}(\omega)$ introduced in \eqref{eq:ac-sus-def} and the small perturbing Hamiltonian part from \eqref{eq:h-tot} renders the usual Zeeman interaction $\delta H_\mathrm{pert}(t) = -\mu B$ (as mentioned before, we approximate $B \approx \mathrm{\mu_0}H$ with $\mathrm{\mu_0}=1$ for dilute samples). An explicit formulation of \eqref{eq:kubo-sus}, together with an alternative derivation based on time-dependent perturbation theory, connection to Green's functions and the Dyson/Volterra series, \cite{Mahan2000,Dyson1949, Volterra1959,Gripenberg1990} is given in Section S3 of the SI.

{\bf Linear response for open quantum systems.}
To model spin-phonon-driven relaxation, we now treat the magnetic center, i.e. the molecular spin, as an open quantum system $S$ coupled to a bosonic thermal bath $B$ of phonons. The unperturbed Hamiltonian $H_0$ in the absence of the external driving part now combines both the system, the bath, as well as the interaction between them \cite{BreuerPetruccione2002}
\begin{equation}\label{eq:h-0}
    H_0 = H_\mathrm{S} \otimes I_\mathrm{B} + I_\mathrm{S} \otimes H_\mathrm{B} + H_\mathrm{I},
\end{equation}
while the perturbation in the weak coupling regime is assumed to act only on the system degrees of freedom,
\begin{equation}
    \delta H_\mathrm{pert}(t) = -F_\mathrm{A}(t)(A_\mathrm{S}\otimes I_\mathrm{B}) \equiv -F_\mathrm{A}(t) A.
\end{equation}
We are interested in the response of an observable $X = X_\mathrm{S} \otimes I_\mathrm{B}$ connected only to the spin system of interest. Assuming that the composite system is initially in thermal equilibrium with respect to $H_0$, i.e. $\rho_\mathrm{eq} = e^{-\beta H_0}/\mathrm{Tr}(e^{-\beta H_0})$, the exact response function can now be written as an analog of \eqref{eq:kubo-response}
\begin{equation}
    \phi_\mathrm{AX}(t) = \frac{i}{\mathrm{\hbar}}\mathrm{Tr}_\mathrm{S+B}\left(Xe^{-i\mathcal{L}_0 t}\{[A,\rho_\mathrm{eq}]\}\right),
\end{equation}
with the new Liouvillian $\mathcal{L}_0 O = [H_0,O]/\mathrm{\hbar}$ from \eqref{eq:h-0} and the trace taken over the combined Hilbert space of the system $S$ and the bath $B$.

Now, we want to find an effective open system equation for $\phi_\mathrm{AX}(t)$ under the influence of bath $B$ that only involves the system $S$ degrees of freedom. Building on the contribution of Uchiyama et al.\cite{Uchiyama2009}, we introduce an operator that lives in the Hilbert space of the system $S$ only
\begin{equation}\label{eq:rrdo-def}
  \rho_A(t) = \mathrm{Tr}_\mathrm{B}\left(e^{-i\mathcal{L}_0 t}\{[A,\rho_{eq}]\}\right), 
\end{equation}
and call it the reduced response density operator (RRDO). As it is not a true density operator (e.g. it is traceless and not normalized), it resembles one because the desired response function is given in an expectation value form for the observable $X$ with respect to $\rho_\mathrm{A}(t)$ and only involves the spin system degrees of freedom similar to the true reduced density of an open quantum system
\begin{equation}\label{eq:open-response-rrdo}
    \phi_\mathrm{AX}(t) = \frac{i}{\mathrm{\hbar}}\mathrm{Tr}_\mathrm{S}\bigl(X\rho_\mathrm{A}(t)\bigr).
\end{equation}
Moreover, using the Nakajima-Zwanzig (NZ) projection-operator formalism, \cite{Nakajima1958,Zwanzig1960} following \cite{Uchiyama2009} we show in Section S4 of the SI that $\rho_\mathrm{A}(t)$ obeys an exact non-Markovian integrodifferential equation of motion analogous to the true reduced density, i.e.
\begin{equation}\label{eq:rrdo-diff-eq}
    \frac{\partial}{\partial t}\rho_\mathrm{A}(t) = -i L\rho_\mathrm{A}(t) + \psi(t) - \int_0^t \mathcal{K}_\mathrm{r}(t-s)\rho_\mathrm{A}(s)ds,
\end{equation}
where $L$ is the projected Liouvillian (unitary evolution part under the stationary bath assumption), $\mathcal{K}_\mathrm{r}(t-s)$ is the response memory kernel, and $\psi(t)$ is an inhomogeneous term encoding initial system-bath correlations and their evolution. \cite{Shibata1977,Weiss2012,Rivas2012,Vega2017}

To connect the exact RRDO equation of motion \eqref{eq:rrdo-diff-eq} with the experimentally relevant complex a.c. susceptibility, the key next step is to reformulate the problem in the frequency domain to obtain a one-sided Fourier transform of RRDO as dictated by the susceptibility to response function relation from \eqref{eq:kubo-sus}. This is only possible because, for a time-independent $H_0$, the unperturbed coupled system-bath problem is stationary, so the memory kernel depends only on the time difference $t-s$. As a result, \eqref{eq:rrdo-diff-eq} has a convolution structure, which is essential because convolutions become simple products after Fourier-Laplace transformation. In parallel, the initial RRDO is given by $\rho_\mathrm{A}(0)=\mathrm{Tr}_\mathrm{B}([A,\rho_\mathrm{eq}])$, where $\rho_\mathrm{eq}$ is defined with respect to the full unperturbed Hamiltonian \eqref{eq:h-0}, including the system-bath coupling $H_\mathrm{I}$. Thus, the response is formulated around the true thermal equilibrium of the interacting system rather than an artificially factorized state. This has later profound importance for physically correct static and dynamical susceptibility results obtained from the simulations, as system-bath correlations contribute both to the initial condition and to the inhomogeneous term of the response equation. In this regard, we transform the RRDO equation \eqref{eq:rrdo-def} from the time domain to the frequency domain. Defining the one-sided Fourier-Laplace transform (FLT) of an operator $O$; $\hat{O}(\omega) = \int_0^{\infty}e^{i\omega t}O(t)dt$ and imposing the condition of dissipative decay of system–bath correlations for the continuous environment spectrum (or, alternatively, adiabatic switching $e^{-\eta t}$ included in the FLT) ensuring that the boundary term after Fourier-transforming \eqref{eq:rrdo-diff-eq} vanishes (see (S4.1.17) for the exact steps) leads to
\begin{equation}\label{eq:rhoA-operator-form}
  \hat{\rho}_\mathrm{A}(\omega) = \Big[iL + \hat{\mathcal{K}}_\mathrm{r}(\omega) - i\omega \mathrm{I}\Big]^{-1} \Big(\rho_\mathrm{A}(0) + \hat{\psi}(\omega)\Big),
\end{equation}
where $\hat{\mathcal{K}}_\mathrm{r}(\omega)$ and $\hat{\psi}(\omega)$ are the Fourier-Laplace transforms of $\mathcal{K}_\mathrm{r}(t)$ and $\psi(t)$, respectively. Then, combining \eqref{eq:sus-omega-heaviside} with \eqref{eq:open-response-rrdo} gives the frequency-dependent susceptibility of the open quantum system by the following formula \cite{Uchiyama2009}
\begin{equation}\label{eq:final-open-sus-main}
    \begin{aligned}
    &\chi_\mathrm{AX}(\omega) = \hat{\phi}_\mathrm{AX}(\omega) \\&= \frac{i}{\mathrm{\hbar}}\mathrm{Tr}_\mathrm{S}\left(X\Big[iL + \hat{\mathcal{K}}_\mathrm{r}(\omega) -i\omega I\Big]^{-1}\big(\rho_\mathrm{A}(0) + \hat{\psi}(\omega)\big)\right).
    \end{aligned}
\end{equation}
It is possible to show (see Section S4.5) that in the absence of system-bath coupling, both $\hat{\mathcal{K}}_\mathrm{r}(\omega)$ and $\hat{\psi}(\omega)$ vanish, and \eqref{eq:final-open-sus-main} reduces to the closed-system Kubo expression \eqref{eq:kubo-sus}.

For the numerical simulations, we wish to vectorize operators and work in the Liouville (Hilbert-Schmidt) space. Assuming the Hilbert space of the spin system $S$ is N-dimensional, we can write all density operators together with $\rho_\mathrm{A}(t)$ as column vectors of length $N^2$ and represent superoperators as $N^2\times N^2$ matrices. \cite{Fano1957,Blum2012} Then \eqref{eq:final-open-sus-main} can be rewritten as
\begin{equation}\label{eq:final-response-matrix-form}
    \chi_\mathrm{AX}(\omega) = \frac{i}{\mathrm{\hbar}}\mathrm{Tr}_\mathrm{S}\left(X(\Xi_\mathrm{R}(\omega)\rho_\mathrm{S}^\mathrm{eq})\right),
\end{equation}
where $\rho_\mathrm{S}^\mathrm{eq} = e^{-\beta H_\mathrm{S}}/\mathrm{Tr}_\mathrm{S}(e^{-\beta H_\mathrm{S}})$ and
\begin{equation}\label{eq:rr-super-op}
    \begin{aligned}
    \Xi_\mathrm{R}(\omega) &= \left[\frac{i}{\mathrm{\hbar}}M_\mathrm{L} + \frac{1}{\mathrm{\hbar}^2}\hat{M}_{\mathcal{K}_\mathrm{r}}(\omega) - i\omega \mathrm{I}\right]^{-1} \times \\ & \times \left(M_{\rho_\mathrm{A}(0)} + \frac{i}{\mathrm{\hbar}}\hat{M}_\mathrm{\psi}(\omega)\right)    
    \end{aligned}
\end{equation}
is the reduced response superoperator in the Liouville space. Therefore, $\Xi_\mathrm{R}(\omega)$ is given as a function of four superoperator matrices: the projected Liouvillian $M_\mathrm{L}$, the Fourier-Laplace transforms of the memory-kernel $\hat{M}_{\mathcal{K}_\mathrm{r}}(\omega)$, and the inhomogeneous-term $\hat{M}_\mathrm{\psi}(\omega)$, together with the matrix $M_{\rho_\mathrm{A}(0)}$ representing the initially correlated RRDO. The first matrix corresponding to the Liouvillian $L$ in the superoperator space is simply given as
\begin{equation}\label{eq:ml}
    (M_\mathrm{L})_{ab, cd} = (H_\mathrm{S})_{a,c}\delta_{d,b}-\delta_{a,c}(H_\mathrm{S})_{d,b},
\end{equation}
while explicit expressions for $\hat{M}_{\mathcal{K}_\mathrm{r}}(\omega)$, $\hat{M}_\mathrm{\psi}(\omega)$, and $M_{\rho_\mathrm{A}(0)}$ are given and derived in Sections S4.1--S4.4 and S4.6. Physically, $\hat{M}_{\mathcal{K}_\mathrm{r}}(\omega)$ block is the central dissipative object of the theory, as it encodes the phonon-assisted transitions and dephasing processes responsible for magnetic relaxation, while $M_\mathrm{L}$ determines the purely coherent oscillatory behavior of the response induced by the unitary evolution ensuring the reduction to the closed-system Kubo formula in the absence of dissipation. Furthermore, $\hat{M}_\mathrm{\psi}(\omega)$ is an inhomogeneous boundary part that occurs because the response is built around the true interacting equilibrium rather than an uncorrelated Born product state, which accompanies the evolution of the term $M_{\rho_\mathrm{A}(0)}$ that encodes how the perturbation operator $A$ initializes the response carrying at the same time the information on the specific observable-input relation in the experiment. Next, we will proceed to give explicit second-order (with respect to the system-bath interaction $H_\mathrm{I}$) expressions for the above-mentioned matrices, considering the harmonic phonon bath and the linear vibronic spin-phonon coupling.

{\bf Second-order explicit expressions for linear spin-phonon coupling.}
To specialize the open-system response formalism to the case of a harmonic phonon bath linearly coupled to the spin system (linear vibronic coupling - LVC) we consider the unperturbed Hamiltonian from \eqref{eq:h-0} where now $H_\mathrm{S}$ is the ab initio Hamiltonian that is to be taken directly from quantum-chemical calculations, $H_\mathrm{B}$ is the harmonic phonon Hamiltonian, and $H_\mathrm{I}$ is the spin-phonon interaction.
\paragraph{Harmonic phonon bath.}
The lattice dynamics are treated within the standard harmonic approximation \cite{Ziman1972,Ashcroft1976,Califano1981,Dove1993,Togo2015,Kittel2018,Srivastava2022}, and since we implement all related computational procedures ourselves, Appendix G in the SI contains a comprehensive coverage of the necessary theory. In terms of phonon normal modes labeled by wavevector $\mathbf{q}$ and band index $j$, with frequencies $\omega_{\mathbf{q},j}$, the bath Hamiltonian reads
\begin{equation}\label{eq:hb-phonons}
    H_\mathrm{B} = \sum_{\mathbf{q},j}\mathrm{\hbar}\omega_{\mathbf{q},j}\left(n_{\mathbf{q},j} + \frac{1}{2}\right),
\end{equation}
with bosonic phonon number operators $n_{\mathbf{q},j}=a^{\dagger}_{\mathbf{q},j}a_{\mathbf{q},j}$ and ladder operators $a_{\mathbf{q},j},a^{\dagger}_{\mathbf{q},j}$ obeying the well-known commutation rules $[a_{\mathbf{q},j},a^{\dagger}_{\mathbf{q'},j'}]=\delta_{\mathbf{q}\mathbf{q'}}\delta_{jj'}$.
The corresponding canonical normal coordinates $Q_{\mathbf{q},j}$ are related to the ladder operators by
\begin{equation}\label{eq:normal-coord}
  Q_{\mathbf{q}, j} = \sqrt{\frac{\mathrm{\hbar}}{2\omega_{\mathbf{q},j}}}\left(a^{\dagger}_{-\mathbf{q},j}+a_{\mathbf{q},j}\right).
\end{equation}
We assume that the bath is in thermal equilibrium, thus neglecting any influence caused by the spin system, which is consistent with the assumption of an infinitely large reservoir that is able to immediately equilibrate and dissipate excess heat. \cite{Albino2021} We take $\rho_\mathrm{B}^\mathrm{eq} = e^{-\beta H_\mathrm{B}}/\mathrm{Tr}_\mathrm{B}(e^{-\beta H_\mathrm{B}})$. In this case, the phonon occupations follow the Bose-Einstein statistics
\begin{equation}
    \bar{n}_{\mathbf{q},j} = \braket{n_{\mathbf{q},j}}_\mathrm{B} = \frac{1}{e^{\beta\mathrm{\hbar}\omega_{\mathbf{q},j}}-1},
\end{equation}
with $\beta=1/(\mathrm{\mathrm{k_B}}T)$. As a consequence of these assumptions, our model will not apply to the description of non-equilibrium effects such as the phonon bottleneck. \cite{Tesi2016}
\paragraph{Linear vibronic (LVC) expansion of the spin Hamiltonian.} The ab initio system $S$ Hamiltonian $H_\mathrm{S}$ corresponding to a given chemical entity depends parametrically on the nuclear coordinates, or equivalently on the Cartesian displacements $\mu_{\alpha}(nl)$ of atom $n$ in unit cell $l$ along Cartesian direction $\alpha$.  In the harmonic regime, we can expand $H_\mathrm{S}$ around the relaxed equilibrium geometry $\{\mu_{\alpha}(nl)\}=0$,
\begin{equation}\label{eq:hs-cartesian-expansion}
    H_\mathrm{S}(\mu) = H_\mathrm{S}^{(0)} + \sum_{n,l,\alpha}\left.\frac{\partial H_\mathrm{S}}{\partial\mu_{\alpha}(nl)}\right|_{0}\mu_{\alpha}(nl)+ \mathcal{O}(\mu^2),
\end{equation}
where $H_\mathrm{S}^{(0)}\equiv H_\mathrm{S}(\{\mu\}=0)\equiv H_\mathrm{S}$ is the spin Hamiltonian at the relaxed structure (it represents system $S$ Hamiltonian in \eqref{eq:h-0}), and the higher-order ($\mathcal O(\mu^2)$) terms correspond to quadratic and anharmonic vibronic couplings which are neglected in the present LVC treatment. The displacements can be expressed in terms of phonon normal coordinates using a mass-weighted expression (see (G.2.54))
\begin{equation}\label{eq:cart-normal-coordinates}
    \mu_{\alpha}(nl) = \frac{1}{\sqrt{L M_n}} \sum_{\mathbf{q},j}e_{\mathbf{q},j}(n\alpha)Q_{\mathbf{q},j}e^{i\mathbf q\cdot\mathbf R_l},
\end{equation}
where $L$ is the number of unit cells, $M_n$ is the mass of the atom $n$, $e_{\mathbf{q},j}(n\alpha)$ are the normalized phonon eigenvectors obtained from diagonalization of the dynamical matrix (see (G.2.34)), and $\mathbf R_l$ is the lattice vector of cell $l$. Substituting \eqref{eq:cart-normal-coordinates} into \eqref{eq:hs-cartesian-expansion} and collecting terms proportional to each normal coordinate $Q_{\mathbf{q},j}$ gives the linear vibronic expansion
\begin{equation}\label{eq:hs-linear-normal-mode}
    H_\mathrm{S}(Q) = H_\mathrm{S}^{(0)} + \sum_{\mathbf{q},j} Y^{\mathbf{q},j} Q_{\mathbf{q},j} + \mathcal{O}(Q^2),
\end{equation}
where the spin-phonon coupling operators $Y^{\mathbf{q},j}$ are defined as directional derivatives of $H_\mathrm{S}$ along the normal modes and can be given as a function of Cartesian derivatives,
\begin{equation}\label{eq:spin-phonon-normal-cart}
    Y^{\mathbf{q},j} = \frac{1}{\sqrt{L}} \sum_{n,l,\alpha}\frac{e_{\mathbf{q},j}(n\alpha)}{\sqrt{M_n}}e^{i\mathbf{q}\cdot\mathbf{R}_l}\left.\frac{\partial H_\mathrm{S}}{\partial\mu_{\alpha}(nl)}\right|_{0}.
\end{equation}
In practice, the derivatives $\partial H_\mathrm{S}/\partial \mu_{\alpha}(nl)$ will be obtained from the finite-difference, numerical, ab initio gradient of the total molecular Hamiltonian with respect to Cartesian displacements and then projected onto the phonon eigenvectors \cite{Mariano2023} directly without using a popular method of expressing them as crystal-field parameters (CFP - Stevens' operators) \cite{Stevens1952,Orbach1961,Shrivastava1983,Chibotaru2012} as done in the majority of studies (see Section S5).

In this framework, the system-bath interaction Hamiltonian from \eqref{eq:h-0} can thus be written as up to linear order in $Q_{\mathbf{q},j}$ as
\begin{equation}\label{eq:hi-lvc}
    H_\mathrm{I} = \sum_{\mathbf{q},j}Y^{\mathbf{q},j}\otimes Q_{\mathbf{q},j},
\end{equation}
where the summation runs over all wave vectors $\mathbf{q}$ in the Brillouin zone and phonon bands $j$, while $H_\mathrm{S}\equiv H_\mathrm{S}^{(0)}$ from \eqref{eq:hs-linear-normal-mode} for the relaxed geometry and will be used as the unperturbed system $S$ Hamiltonian in the projected Liouvillian $L$. Equation \eqref{eq:hi-lvc} is the standard linear vibronic (LVC) coupling adopted in the rest of this work.

\paragraph{Explicit second-order expressions for the Liouville-space superoperators entering $\Xi_\mathrm{R}(\omega)$} (for the extended derivations corresponding to this paragraph, refer to Section S4.7 in the SI). It is now the most convenient to work in the eigenbasis of $H_\mathrm{S}$,
\begin{equation}
  H_\mathrm{S} \ket{a} = \mathrm{\hbar}\omega_a \ket{a},
\end{equation}
and introduce the transition frequencies
\begin{equation}\label{eq:omega-ab}
  \omega_{ab} = \omega_a - \omega_b,
\end{equation}
which correspond to energy differences between the states $\ket{a}$ and $\ket{b}$ of our future ab initio molecular Hamiltonian. All the following formulas will be given as component equations in this basis, e.g.
\begin{equation}
  (Y^{\mathbf{q},j})_{a,b} = \bra{a}Y^{\mathbf{q},j}\ket{b},\qquad (H_\mathrm{S})_{a,b} = \mathrm{\hbar}\omega_a\delta_{a,b},
\end{equation}
while a generic superoperator $\mathcal{S}$ is represented by a doubly-indexed matrix with components
\begin{equation}
  S_{ab,cd} = \bra{a}\mathcal{S}\big(\ket{c}\bra{d}\big)\ket{b},
\end{equation}
where we used $\{\ket{a}\bra{b}\}$, built from the eigenstates of $H_\mathrm{S}$, as an operator basis. Further, for the linear interaction in the form of \eqref{eq:hi-lvc}, all dependence on the phonon bath $B$ in the kernels expanded to the second order in perturbation is contained in two-time phonon bath correlation functions \cite{Legget1987}
\begin{equation}\label{eq:bath-corr}
    \begin{aligned}
    &\big\langle Q_{\mathbf{q},j}(t)Q_{\mathbf{-q},j}(t')\big\rangle_\mathrm{B} \\&= \frac{\mathrm{\hbar}}{2\omega_{\mathbf{q},j}}\left[\bar{n}_{\mathbf{q},j}e^{i\omega_{\mathbf{q},j}(t-t')} + (\bar{n}_{\mathbf{q},j}+1)e^{-i\omega_{\mathbf{q},j}(t-t')}\right], 
    \end{aligned}
\end{equation}
which give rise to two types of matrix-valued phonon spectral densities
\begin{equation}\label{eq:j}
    \begin{aligned}
    J^{\mathbf{q},j}_{a,b}(\omega) &= \frac{\mathrm{\hbar}\mathrm{\pi}}{\omega_{\mathbf{q},j}} \Big[\bar{n}_{\mathbf{q},j}\delta(\omega_{ab}-\omega_{\mathbf{q},j}-\omega)\\& + (\bar{n}_{\mathbf{q},j}+1)\delta(\omega_{ab}+\omega_{\mathbf{q},j}-\omega)\Big],
\end{aligned}
\end{equation}
and the correlated analog,
\begin{equation}\label{eq:j-corr}
    \begin{aligned}
    &\mathcal{J}^{\mathbf{q},j}_{a,b}(\omega,\omega_{cd})\\&= \frac{\mathrm{\hbar\pi}}{\omega_{\mathbf{q},j}}\Big[\bar{n}_{\mathbf{q},j}\zeta(\omega_{ab}+\omega_{\mathbf{q},j})\delta(\omega_{cd}+\omega_{ab}+\omega_{\mathbf{q},j}-\omega)\\&+ (\bar{n}_{\mathbf{q},j}+1)\zeta(\omega_{ab}-\omega_{\mathbf{q},j})\delta(\omega_{cd}+\omega_{ab}-\omega_{\mathbf{q},j}-\omega)\Big],
    \end{aligned}
\end{equation}
where $\delta(\omega)$ is the Dirac delta function and
\begin{equation}\label{eq:dzeta}
  \zeta(\omega) = \int_0^\beta e^{\mathrm{\hbar}\omega s}ds =
  \begin{cases}
    \dfrac{e^{\mathrm{\hbar}\beta\omega}-1}{\mathrm{\hbar}\omega}, & \omega\neq 0, \\[6pt]
    \beta, & \omega = 0.
  \end{cases}
\end{equation}
The $\delta$ functions are a consequence of harmonic approximation for lattice dynamics, which yields infinitely narrow linewidths and infinitely long phonon relaxation rates for their undamped motion. In practice, they need to be replaced by broadened lineshapes: Lorentzians $\delta_\mathrm{L}(\omega)$,
\begin{equation}\label{eq:delta-l}
    \delta(\omega) \to \delta_\mathrm{L}(\omega) = \frac{1}{\mathrm{\pi}}\frac{\Delta}{\omega^2 + \Delta^2},
\end{equation}
or Gaussians $\delta_\mathrm{G}(\omega)$
\begin{equation}\label{eq:delta-g}
    \delta(\omega) \to \delta_\mathrm{G}(\omega) =\frac{1}{\sigma\sqrt{2\mathrm{\pi}}} e^{-\omega^2/2\sigma^2},
\end{equation}
with a finite FWHM $\Gamma$ related to the above parameters by $\Delta = \Gamma /2$ and $\sigma = \Gamma /(2\sqrt{2ln2})$, as done and discussed in \cite{Lunghi2022}, to enable the numerical treatment and at the same time to mimic finite phonon relaxation times present in real crystal lattices in the presence of higher order (anharmonic) terms \cite{Lunghi2017_1,Nabi2023} and scattering (see the discussion following (S4.7.19)). The numerical procedure of converging grid density in the Brillouin-zone grid vs $\Gamma$ value will be discussed in the Computational Methods section. The corresponding Fourier-Laplace transforms needed for the final expressions are denoted as $\hat{J}^{\mathbf{q},j}_{a,b}(\omega)$ and $\hat{\mathcal{J}}^{\mathbf{q},j}_{a,b}(\omega)$. They can be written generically as
\begin{equation}
  \hat{J}^{\mathbf{q},j}_{a,b}(\omega) = \frac{1}{2}J^{\mathbf{q},j}_{a,b}(\omega) + \frac{i}{2}\,\mathcal{H}[J^{\mathbf{q},j}_{a,b}](\omega),
\end{equation}
where $\mathcal{H}$ is the Hilbert transform (see Appendix A8) with respect to $\omega$. To obtain the explicit expression for broadened (finite FWHM) versions, one needs to replace $\delta$-functions in \eqref{eq:j} and \eqref{eq:j-corr} with the one-sided Fourier-Laplace transforms of \eqref{eq:delta-l} and \eqref{eq:delta-g} for Lorentzian and Gaussian lineshapes, respectively. The latter read
\begin{equation}\label{eq:delta-l-flp}
    \hat{\delta}_\mathrm{L}(\omega) =  \frac{1}{2\mathrm{\pi}}\left(\frac{\Delta}{\omega^2 + \Delta^2}+i\frac{\omega}{\omega^2 + \Delta^2}\right),
\end{equation}
and
\begin{equation}\label{eq:delta-g-flp}
    \hat{\delta}_\mathrm{G}(\omega) =\frac{1}{2\sigma\sqrt{2\mathrm{\pi}}} \left( e^{-\omega^2/2\sigma^2} +i\frac{2}{\sqrt{\mathrm{\pi}}}\mathcal{D}\left(\frac{\omega}{\sqrt{2}\sigma}\right)\right),
\end{equation}
where $\mathcal{D}(\omega)$ is the Dawson function defined in the SI (A.9.2).
Lastly, we need one more scalar function, which also appears later in connection with imaginary time Dyson integrals
\begin{equation}\label{eq:int}
  I(\omega_1,\omega_2) = \int_0^\beta d\tau_1 \int_0^{\tau_1} d\tau_2 e^{\mathrm{\hbar}(\omega_1\tau_1+\omega_2\tau_2)},
\end{equation}
for which closed analytic forms (including all special cases) are listed in (S4.7.29).

Having set the notation, in points (i)-(iv), we summarize the explicit second-order expressions for the Liouville-space blocks entering $\Xi_\mathrm{R}(\omega)$ in \eqref{eq:rr-super-op} for the harmonic phonon bath and LVC coupling. Detailed step-by-step derivations of these can be found in Section S4.7 of the SI.

\noindent{\it (i) Homogeneous projected Liouvillian block $M_\mathrm{L}$.}
In the weak-coupling projector scheme with $\mathcal{P}O = \mathrm{Tr}_\mathrm{B}(O)\otimes\rho_\mathrm{B}^{\mathrm{eq}}$ and vanishing odd bath moments, the projected Liouvillian reduces to the system Liouvillian $L=\mathcal L_\mathrm{S}$, and in the eigenbasis of $H_\mathrm{S}$ we have
\begin{equation}\label{eq:ml}
  (M_\mathrm{L})_{ab,cd} = (H_\mathrm{S})_{a,c}\delta_{d,b} - \delta_{a,c}(H_\mathrm{S})_{d,b},
\end{equation}
or, in compact Kronecker form,
\begin{equation}
  M_\mathrm{L} = H_\mathrm{S}\otimes \mathrm{I} - \mathrm{I}\otimes H_\mathrm{S}^\mathrm{T}.
\end{equation}

\noindent{\it (ii) Fourier-Laplace transformed memory kernel block $\hat{M}_{\mathcal{K}_\mathrm{r}}(\omega)$.} Inserting \eqref{eq:hi-lvc} into the second-order convolution kernel, using the bath correlators from \eqref{eq:bath-corr} and performing the one-sided Fourier-Laplace transform gives
\begin{equation}\label{eq:mk-hat}
    \begin{aligned}
     &(\hat{M}_{\mathcal{K}_\mathrm{r}}(\omega))_{ab,cd}\\&= \sum_{\mathbf{q},j}\Big\{\sum_e\hat{J}^{\mathbf{q},j}_{e,d}(\omega)Y^{\mathbf{q},j}_{a,e} Y^{-\mathbf{q},j}_{e,c}\delta_{d,b} - \hat{J}^{\mathbf{q},j}_{a,d}(\omega)Y^{\mathbf{q},j}_{a,c} Y^{-\mathbf{q},j}_{d,b}\\&+ \sum_e\hat{J}^{\mathbf{q},j}_{e,c}(-\omega)\delta_{a,c}Y^{\mathbf{q},j}_{d,e} Y^{-\mathbf{q},j}_{e,b} - \hat{J}^{\mathbf{q},j}_{b,c}(-\omega)Y^{\mathbf{q},j}_{a,c} Y^{-\mathbf{q},j}_{d,b}\Big\},   
    \end{aligned}
\end{equation}
where we note that the lattice symmetry implies $Y^{-\mathbf{q},j}_{a,b}=(Y^{\mathbf{q},j}_{b,a})^{*}$. As we earlier connected this block with dissipative relaxation, we note here that its explicit dependence on the probing input frequency $\omega$ reflects the fact that the system-bath response is not instantaneous; therefore, it encodes non-Markovian relaxation with memory effects.

\noindent{\it (iii) Inhomogeneous block \(\hat{M}_\mathrm{\psi}(\omega)\).} Initial system-bath correlations lead to an inhomogeneous contribution $\psi(t)$ whose Fourier-Laplace transform defines $\hat{M}_\mathrm{\psi}(\omega)$. This can be viewed as an additional source term to the response contributed by pre-existing correlations at the time of the perturbation application. At second order in $H_\mathrm{I}$ these can be conveniently expressed through correlated spectral densities $\mathcal{J}^{\mathbf{q},j}$ as
\begin{equation}\label{eq:mpsi-hat}
    \begin{aligned}
    (\hat{M}_\mathrm{\psi}(\omega))_{ab,cd} &= \sum_{\mathbf{q},j}\sum_e\Big\{\hat{\mathcal{J}}^{\mathbf{q},j}_{d,b}(\omega,\omega_{ec})Y^{\mathbf{q},j}_{a,e} A_{e,c} Y^{-\mathbf{q},j}_{d,b}\\& - \hat{\mathcal{J}}^{\mathbf{q},j}_{d,e}(\omega,\omega_{ac})A_{a,c} Y^{\mathbf{q},j}_{d,e} Y^{-\mathbf{q},j}_{e,b}\\& + \sum_f \hat{\mathcal{J}}^{\mathbf{q},j}_{d,e}(\omega,\omega_{ef})\delta_{a,c} Y^{\mathbf{q},j}_{d,e} A_{e,f} Y^{-\mathbf{q},j}_{f,b}\\& - \hat{\mathcal{J}}^{\mathbf{q},j}_{d,e}(\omega,\omega_{eb})Y^{\mathbf{q},j}_{a,c} Y^{-\mathbf{q},j}_{d,e} A_{e,b}\Big\}.  
    \end{aligned}
\end{equation}
Neglecting equilibrium system-bath correlations amounts to setting $\hat{M}_\mathrm{\psi}(\omega)\approx 0$ and the results of such an approximation are explored numerically in subsequent parts of the article.

\noindent{\it (iv) Initial RRDO block $M_{\rho_\mathrm{A}(0)}$.} The initial reduced response density operator $\rho_\mathrm{A}(0)=\mathrm{Tr}_\mathrm{B}([A,\rho_\mathrm{eq}])$ receives a second-order correction from $H_\mathrm{I}$ due to the correlated initial bath density $\rho_\mathrm{eq}$. This can be encoded as an action of a superoperator matrix on the density operator of system $S$, $\rho_\mathrm{A}(0) = M_{\rho_\mathrm{A}(0)}\rho_\mathrm{S}^\mathrm{eq},$ where the components are given by
\begin{equation}\label{eq:m-rho-a}
    \begin{aligned}
    &(M_{\rho_\mathrm{A}(0)})_{ab,cd} = A_{a,c}\delta_{d,b} - \delta_{a,c}A_{d,b} \\& - \sum_{\mathbf{q},j}\frac{\mathrm{\hbar}}{2\omega_{\mathbf{q},j}} \sum_e \Big\{\big[\bar{n}_{\mathbf{q},j}I(\omega_{de}+\omega_{\mathbf{q},j},\omega_{eb}-\omega_{\mathbf{q},j})\\& + (\bar{n}_{\mathbf{q},j}+1) I(\omega_{de}-\omega_{\mathbf{q},j}, \omega_{eb}+\omega_{\mathbf{q},j})\big] A_{a,c} Y^{\mathbf{q},j}_{d,e} Y^{-\mathbf{q},j}_{e,b}\\& - \sum_f \big[ \bar{n}_{\mathbf{q},j} I(\omega_{de}+\omega_{\mathbf{q},j}, \omega_{ef}-\omega_{\mathbf{q},j})\\& + (\bar{n}_{\mathbf{q},j}+1)I(\omega_{de}-\omega_{\mathbf{q},j},\omega_{ef}+\omega_{\mathbf{q},j})\big]\delta_{a,c} Y^{\mathbf{q},j}_{d,e} Y^{-\mathbf{q},j}_{e,f} A_{f,b}\Big\}.   
    \end{aligned}
\end{equation}
In the simplest (uncorrelated) initial equilibrium approximation, where the equilibrium density matrix factorizes as $\rho_\mathrm{eq} \approx \rho_\mathrm{S} \otimes \rho_\mathrm{B}$, one retains only the first two terms, $M_{\rho_\mathrm{A}(0)}\approx A\otimes \mathrm{I} - \mathrm{I} \otimes A^T$, while \eqref{eq:m-rho-a} adds the leading correction from equilibrium spin-phonon correlations. An alternative, numerically simple extension is also discussed in (S4.6.22-26).

Together, eqs. \eqref{eq:final-response-matrix-form}, \eqref{eq:rr-super-op} for $\chi_\mathrm{AX}(\omega)$ and $\Xi_\mathrm{R}(\omega)$ with \eqref{eq:ml}, \eqref{eq:mk-hat}, \eqref{eq:mpsi-hat}, and \eqref{eq:m-rho-a} for superoperator matrices $M_\mathrm{L}$, $\hat{M}_{\mathcal{K}_\mathrm{r}}$, $\hat{M}_\mathrm{\psi}(\omega)$, and $(M_{\rho_\mathrm{A}(0)}$ provide a complete and numerically straightforward recipe for computing the open-system susceptibility $\chi_\mathrm{AX}(\omega)$ for a relevant system $S$ with linear vibronic coupling to a harmonic phonon bath $B$
\begin{equation}\label{eq:final-ac-sus-all}
    \begin{aligned}
    \chi_\mathrm{AX}(\omega) =& \frac{i}{\mathrm{\hbar}}\mathrm{Tr}_\mathrm{S}\left(X\left[\frac{i}{\mathrm{\hbar}}M_\mathrm{L} + \frac{1}{\mathrm{\hbar}^2}\hat{M}_{\mathcal{K}_\mathrm{r}}(\omega) - i\omega \mathrm{I}\right]^{-1}\right. \\ &\left.
    \times \left(M_{\rho_\mathrm{A}(0)} + \frac{i}{\mathrm{\hbar}}\hat{M}_\mathrm{\psi}(\omega)\right)\rho_\mathrm{S}^\mathrm{eq}
    \right).
\end{aligned}
\end{equation}

\section*{Computational Methods}

We now provide the details of the ab initio implementation of the open quantum system formalism just derived. To achieve this, we focus on the three-dimensional cyanido-bridged coordination polymer family \textbf{1}-\textbf{3} in which the lanthanide sites are embedded in a matrix formed by low-spin Co(III) cyanido linkers and diluted by diamagnetic Y(III) centers (see Figure \ref{fig:1}). Previous first-principles treatments have in the majority been restricted to molecular crystals or isolated molecules; \cite{Goodwin2017,Lunghi2017_1,Lunghi2017_2,Lunghi2019,Ho2018,Briganti2021,Mondal2022,Lunghi2022,Kragskow2023,Lunghi2023,Mariano2023} therefore, contrary to those investigations, we expect a much more pronounced impact on the magnetic relaxation caused by the phonon dispersion within the Brillouin zone, along with the strong role of acoustic branches. In this regard, we outline here, step by step, the computational scheme utilized.

{\bf Periodic density functional theory (pDFT) calculations and phonon simulations.}
We started with the experimental structure of the undiluted analog of compound \textbf{2} - \( \{ \mathrm{Tb}^{\text{III}}[ \mathrm{Co}^{\text{III}}(\mathrm{CN})_6 ]\}\), which was recently published in \cite{Wang2023} as a rare example of an SMM-based luminescent thermometer. Since we are only interested in samples that are strongly magnetically diluted with Y$^{\text{3+}}$ ions and contain only a few percent of the magnetic lanthanide centers, we propose to assume that all investigated systems \textbf{1}-\textbf{3 }approximately acquire the crystal structure of the pure \( \{ \mathrm{Y}^{\text{III}}[ \mathrm{Co}^{\text{III}}(\mathrm{CN})_6 ]\}\) diamagnetic phase (\textbf{4}); therefore, we substituted Tb atoms with Y in the crystal structure from X-ray diffraction and conducted geometry optimization with the pDFT method implemented in the CP2K version 2024.3 software. \cite{Kuhne2020} For the relaxation and geometry optimization of the unit cell containing 28 atoms (see Figure \ref{fig:1}), we used very tight convergence criteria of $1\cdot10^{-7}$ a.u. for gradients (forces) and $1\cdot10^{-9}$ a.u. for SCF energies, as those results must be further suitable for numerical differentiation using the finite difference method. We employed the PBE exchange-correlation functional \cite{Perdew1996} with a suitable TZVP-MOLOPT-PBE-GTH (Goedecker–Teter–Hutter pseudopotentials \cite{Goedecker1996}) basis set for all atoms, setting the plane wave cutoff of 1800 Ry together with a dense integration grid (NGRIDS 5) and DFT-D3 dispersion correction \cite{Grimme2010}. During the entire procedure, we kept the high, initial hexagonal system symmetry constrained. The resultant geometry of the optimized cell is reported in Tables S1 and S2. To serve as a reference, we obtained phase \textbf{4} experimentally and characterized it structurally for the first time (see the Experimental Details Section S6  \cite{Sheldrick2014,Sheldrick2015,Farrugia1999,Casanova2004}) with the results reported in Figures S1, S2, and Tables S3, S4. The comparison of the experimental and optimized structures in Figure S3 and Tables S5, S6 reveals a satisfactory agreement of bond lengths and angles within 0.02 \AA\  and 5$^\circ$ maximal differences, respectively, with the relaxed computational cell having a slightly contracted \textit{c} crystallographic axis, but retaining a volume similar to the experimental one (621.4 and 617.8 \AA$^3$, respectively). To compute the potential energy Hessian matrix required for the phonon structure simulations, we constructed a large enough $3 \times 3 \times 2$ supercell (see Figure S4) as discussed in Appendix G3 and used a finite two-point difference method around the equilibrium geometry by evaluating forces, keeping the very tight convergence criteria, acting on each atom within geometries produced by displacing by $\pm0.01$ \AA\ atoms within the first unit cell. Then the force-constant matrix (Hessian in Cartesian displacements) obtained in this way can be used to construct a dynamical matrix (G.2.34) by mass-weighting and lattice Fourier transform for an arbitrary wavevector of the BZ. Its diagonalization yields normal modes as eigenvectors and eigenvalues as the square of corresponding phonon frequencies $\omega^2$ (see G2.39). We report the $\mathrm{\Gamma}$-point frequencies in Table S7 and superimpose those related to the vibrations of the CN$^-$ bridging ligands on the experimental IR spectra in Figure S5, revealing excellent agreement and validating the employed procedure. Moreover, we also simulated the intensities of the IR absorption discussed in Appendix G4 and presented them in Figure S6. For further reference, we also provide a phonon dispersion plot in Figure S7 along a standard Brillouin zone (BZ) path for hexagonal systems visualized in Figure S8 \cite{Larsen2017} together with the phonon density of states (DOS) in Figure S9 obtained using the special multigrid (discussed later and in Section S7.1) in the BZ covering the entire frequency range. Normal modes computed via the dynamical matrix constructed in this part will serve for the computation of the spin-phonon coupling operator through \eqref{eq:spin-phonon-normal-cart}.

{\bf Ab initio evaluation of spin-phonon coupling operators.}
In the presented implementation, the system Hamiltonian $H_\mathrm{S}$ is not restricted to a phenomenological spin Hamiltonian, but is taken to be the ab initio molecular Hamiltonian obtained from multiconfigurational calculations, including all relevant relativistic interactions.\cite{Mariano2023}  In particular, $H_\mathrm{S}$ corresponds to the effective Hamiltonian assembled as
\begin{equation}\label{eq:model-ham}
  \hat{H}_\mathrm{S} = H_\mathrm{MCH} + H_\mathrm{SOC} + H_\mathrm{SSC} + H_\mathrm{Zeeman},
\end{equation}
where $H_\mathrm{MCH}$ is the molecular Coulomb Hamiltonian, $H_\mathrm{SOC}$ and $H_{SSC}$ describe spin–orbit and spin–spin coupling, and $H_\mathrm{Zeeman} = \mathrm{\mu}_\mathrm{B}(\mathrm{g_e}\hat{\mathbf{S}}+\hat{\mathbf{L}})\cdot \mathbf{H}_\mathrm{d.c.}$ is the Zeeman interaction with the static bias field $\mathbf{H}_\mathrm{d.c.}$ applied along the measurement direction. Thus, the static field used experimentally is already included at the level of $H_\mathrm{S}$ and is fully accounted for in subsequent gradient calculations and open-system dynamics.

The remaining ingredient needed to evaluate the spin–phonon coupling operators $Y^{\mathbf{q},j}$ in \eqref{eq:spin-phonon-normal-cart} is the gradient of the molecular Hamiltonian with respect to Cartesian, nuclear coordinates, $\nabla H_\mathrm{S}$, evaluated at the relaxed geometry ($\mathbf{R}_0$). In Section S5 of the SI, based on \cite{Feynman1939,Taylor2016,Lowdin1950,Plasser2016,Granucci2001,Akimov2018,Mai2015,Pederzoli2017,Lowdin1955}, we show a derivation of a general expression that was implemented and successfully used in \cite{Mariano2023} for studies of magnetic relaxation dynamics beyond the simple spin Hamiltonian formalism
\begin{equation}\label{eq:hamiltonian-derivatives}
    \begin{aligned}
    &\bra{j(\mathbf{R}_0)}\nabla H_\mathrm{S}\ket{i(\mathbf{R}_0)} \\& = \nabla E_i\delta_{ji} + (E_i -E_j)\bigl(K^\mathrm{MCH}_{ji}(\mathbf R_0) + K^\mathrm{U}_{ji}(\mathbf R_0)\bigr),
    \end{aligned}
\end{equation}
where $\ket{i(\mathbf{R}_0)}$ denote the eigenstates of $H_\mathrm{S}$. In Section S5, we also show a highly efficient algorithm used to compute the couplings $K^\mathrm{MCH}$ and $K^\mathrm{U}$ from SA-CASSCF wave functions, exploiting the Schur complement-based determinant formula for the calculation of Slater determinant overlaps. The overall scheme exploits (i) a finite-difference representation of $\ket{\nabla\psi_i}$ in terms of overlaps between relaxed and displaced geometries, (ii) original Schur–complement factorization of CI-determinant overlap matrices tailored to CASSCF active spaces, and (iii) Löwdin-type orthogonalization and phase-tracking procedures that handle near-degeneracies and Kramers pairs in the presence of spin–orbit coupling. As a result, the spin–phonon operators $Y^{\mathbf{q},j}$ entering the response kernel are obtained directly from the fully relativistic ab initio molecular Hamiltonian, without any intermediate mapping to a crystal-field parameter model or effective spin Hamiltonian, contrary to the current standard practice in magnetic relaxation simulations. \cite{Chibotaru2012,Escalera2017} This allows us to treat all low-lying spin–orbit states and excited multiplets, which can contribute to and play a crucial role in magnetization dynamics, on equal footing.

{\bf Multiconfigurational relativistic SA-CASSCF calculations.}
We construct our system's $\mathrm{S}$ model Hamiltonian \eqref{eq:model-ham} at equilibrium geometry for lanthanide ions using operator matrices ($H_\mathrm{MCH}, H_\mathrm{SOC}, L, S$) taken directly from quantum-chemical simulations. To do this, we chose to work with minimal clusters $[\mathrm{Ln}^{\text{III}}(\mathrm{CN})_6 ]^{3-}$ presented in Figure \ref{fig:1} panel b). In this approximation, our cluster only contains the first trigonal prismatic coordination sphere of a lanthanide ion with six CN${^-}$ ligands. The geometry of such a cluster is provided by the pDFT-optimized super-cell geometry hosting diamagnetic phase \textbf{4}, after replacing the Y$^{3+}$ ion with the lanthanides \textbf{1} - Yb$^{3+}$, \textbf{2} - Tb$^{3+}$, and \textbf{3} - Dy$^{3+}$. As we are primarily interested in benchmarking our new methodology against the state-of-the-art of ab initio quantum master equations, \cite{Lunghi2023} this vastly reduces the computational intensity of further investigation as compared to the case where one also would consider, e.g. the whole set of six $[\mathrm{Co}^{\text{III}}(\mathrm{CN})_6 ]^{3-}$ moieties surrounding the magnetic center (39 compared to 237 Cartesian degrees of freedom for numerical differentiation and each normal mode during BZ integration). This can also be justified considering the rapid decay of spin-phonon coupling strength with increasingly distant neighbor atoms from the magnetic ion. The quality of this approximation that neglects the impact of further lying atoms will be assessed later based on the theoretical results compared to the experimental characteristics. We used the SA-CASSCF (State Average Complete Active Space Self-Consistent Field) approach \cite{Roos1980,Werner1981} implemented in ORCA 6.1.0 software package \cite{Neese2025,Neese2023}. To include scalar relativistic effects we employed the two-component second-order Douglas-Kroll-Hess (DKH2) Hamiltonian \cite{Douglas1974,Hess1986,Jansen1989} with the apropriate SARC2-DKH-QZVP basis set \cite{Aravena2016} for lanthanides and minimally augmented variant ma-DKH-def2-SVP \cite{Zheng2010} for the rest of the atoms. All calculations are performed with VeryTight SCF convergence criteria to meet the standards for subsequent numerical differentiation, and we used the RIJCOSX (Resolution-of-the-Identity with Chain of Spheres algorithm) \cite{Neese2003,Neese2009,Helmich2021} with AutoAux (Automatic Generation of Auxiliary Basis Sets) \cite{Stoychev2017} to speed up the computations. The active space was composed of seven 4f-orbitals and 13 active electrons for Yb CAS(13in7), 8 electrons for Tb CAS(8in7), and 9 for Dy CAS(9in7). We optimized all possible roots for the ground spin multiplicity of each lanthanide i.e. doublets, septets, and sextets,  for \textbf{1}, \textbf{2}, and \textbf{3} respectively. Then the spin-free states were mixed by SOC (Spin–Orbit-Coupling) within the AMFI (Atomic Mean-Field) using quasi-degenerate perturbation theory. \cite{Ganyushin2006,Ganyushin2013,Lang2019,Ugandi2023} The resulting energy splittings of the ground term multiplets are reported in Table S8, while we also provide the visualization of the orientation of main magnetic axes of the clusters for the ground energy doublets in Figure S10 as they will later be used as three orthogonal directions for susceptibility averaging. The same type of calculations is performed for $2 \times 3 \times N$ (with $N=13$, number of atoms in the cluster) displaced geometries of the clusters to obtain the Hamiltonian derivatives \eqref{eq:hamiltonian-derivatives} within the above-mentioned process of numerical differentiation with phase-tracking and Schur complement-type algorithm for CASSCF overlap calculations thoroughly explained in Section S5. We note here that in our process, we include the d.c. static bias magnetic field by adding the Zeeman term to the  SOC Hamiltonian produced by including $L$ and $S$ matrices in the SOC basis. Then the Hamiltonian cartesian gradient elements $\nabla H$ can be combined with eigenvectors of the dynamical matrix for a given wavevector to be transformed to normal-mode spin-phonon coupling matrices $Y^{\mathbf{q},j}$ through \eqref{eq:spin-phonon-normal-cart}. As we are using the same Hessian matrix of phase \textbf{4} for all investigated systems at this point, we simply substitute the Y atom mass in the mass-weighted coordinates with a corresponding lanthanide mass to approximate the phonon structure as the local vibrational participation of the Ln site embedded in Y-based matrix.

{\bf Complex a.c. magnetic susceptibility simulations.}
Having set up the machinery for the ab initio evaluation of spin-phonon coupling operators $Y^{\mathbf{q},j}$ for an arbitrary phonon branch and wavevector in BZ, including a static d.c. magnetic field with a given direction, we now discuss the construction of the superoperator matrices entering expression \eqref{eq:rr-super-op}. The $M_\mathrm{L}$ unitary evolution term \eqref{eq:ml} is particularly easy to evaluate, as it is simply a Kronecker product of the ab initio system's Hamiltonian in equilibrium geometry, plus the Zeeman interaction term, as described in \eqref{eq:model-ham}, with the identity matrix, which effectively moves it to the Liouville superoperator space. The next two matrices $\hat{M}_{\mathcal{K}_\mathrm{r}}(\omega)$ \eqref{eq:mk-hat} and $\hat{M}_\mathrm{\psi}(\omega)$ \eqref{eq:mpsi-hat} are explicitly dependent on the magnetic field frequency. Then, since we are interested here in the $\omega$ range between $\approx$ $10^0$ to $10^4$ Hz as those are available in commercial SQUID-type magnetometers, we note that this translates approximately to a range between $3.34\cdot10^{-11}$ and $3.34\cdot10^{-7}$ cm$^{-1}$ in wavenumbers. On the other hand, the smallest splitting of lanthanide energy levels, considered in this contribution, under 3000 Oe d.c. magnetic field has an order of 10$^{-1}$ cm$^{-1}$, while our densest BZ grid used (see the discussion below) has a resolution of around 10$^{-3}$ cm$^{-1}$, considering FWHM used for the Gaussian Broadening with the same threshold of the lowest phonon frequency taken into account. Seeing that the $\omega$ field frequency is at least 4 to 8 orders of magnitude smaller than the considered $\omega_{ab}$ or $\omega_{\mathbf{q},j}$ in \eqref{eq:j} and \eqref{eq:j-corr} and, what is more important, FWHM of the chosen smearing dictating the scale of variation and smoothness of the broadened spectral densities, we are not able to accommodate it within our computational resolution. It means that $\hat{M}_{\mathcal{K}_\mathrm{r}}(\omega)$ and $\hat{M}_\mathrm{\psi}(\omega)$ are effectively constant on this scale and we can approximately set the frequency $\omega$ of the field to zero, as its effect is numerically negligible. We thus approximate these quantities as $\hat{M}_{\mathcal{K}_\mathrm{r}}(0)$ and $\hat{M}_\mathrm{\psi}(0)$, greatly reducing the computational cost of simulations as we now do not need to reconstruct the operators for the hundreds values of frequencies that we are interested in sampling. However, this approximation comes with a few implications. First of all, we will not be able to see any line-shape changes or maximum frequency-shift of imaginary susceptibility because now the boundary terms $\hat{M}_\mathrm{\psi}$ and $M_{\rho_\mathrm{A}(0)}$ are both frequency-independent and act as a uniform multiplier of the resolvent components (they form a source vector in \eqref{eq:rr-super-op} and \eqref{eq:lin-eq-xi}) across the investigated frequency range, giving real and imaginary parts of susceptibility with purely Debye-model-like shapes because they steam now from the also frequency-independent operators $\frac{i}{\mathrm{\hbar}}M_\mathrm{L} + \frac{1}{\mathrm{\hbar}^2}\hat{M}_{\mathcal{K}_\mathrm{r}}$ in the resolvent. Nevertheless, in our current setup, those effects are, as we see from the magnitude of frequency differences, negligible. The second implication of neglecting the magnetic field frequency is that for the selection rules imposed by the Dirac delta functions in \eqref{eq:j} and \eqref{eq:j-corr}. By setting $\omega = 0$ we are left e.g. for \eqref{eq:j} with phonon absorption/emission term $\bar{n}_{\mathbf{q},j}\delta(\omega_{ab}-\omega_{\mathbf{q},j}) +(\bar{n}_{\mathbf{q},j}+1)\delta(\omega_{ab}+\omega_{\mathbf{q},j})$ which is encountered in the standard formulation of the magnetic relaxation problem (see \cite{Lunghi2023} and equations \eqref{eq:one-ph-sec-markow-red}, \eqref{eq:g-one-ph}). It is then customary to treat it as in the previous implementation of the problem (MolForge software \cite{Lunghi2022}) to check the sign of $\omega_{ab}$, and since the $\omega_{\mathbf{q},j}$ frequencies are always positive if $\omega_{ab} > 0$ then only the first phonon-absorption part $\bar{n}_{\mathbf{q},j}\delta(\omega_{ab}-\omega_{\mathbf{q},j})$ contributes, otherwise, if $\omega_{ab} < 0$ we only include the phonon-emission term $(\bar{n}_{\mathbf{q},j}+1)\delta(\omega_{ab}+\omega_{\mathbf{q},j})$. Since, for further numerical integration, the delta functions are replaced with even broadening functions, either Lorentzian or Gaussian, we call this scheme \textit{symmetric}, as phonons with both higher or lower energies than the energy gap $\omega_{ab}$ are included. However, looking at the original form, e.g., of \eqref{eq:j} we must also explore in this contribution an additional energy-conservation-type constraint imposed by the appearance of $-\omega$ part in the delta functions, i.e., the sign of $\omega_{ab}\pm\omega_{\mathbf{q},j}$ must be the same as that of $\omega$ for the argument to be equal to zero. This restriction, even when equating $\omega = 0$ later, has to be preserved because it comes from \eqref{eq:mk-hat} where both $\hat{J}^{\mathbf{q},j}_{a,b}(\omega)$ and $\hat{J}^{\mathbf{q},j}_{a,b}(-\omega)$ are used and must be distinguished. We call this approximation \textit{asymmetrical} since we effectively use only one half of the Lorentzian or Gaussian phonon broadening, depending on the sign of $\omega$, avoiding spurious off-resonant contributions from the otherwise forbidden phonon spectrum region. The second treatment should better reflect the developed non-Markovian linear response framework, in which we see that the simple phonon absorption/emission classification of terms loses meaning in the presence of the a.c. magnetic field. The same reasoning applies to the correlated spectral densities \eqref{eq:j-corr} by looking at the sums $\omega_{cd}+\omega_{ab}$. In the following Results and Discussion section, we report, as a numerical diagnostic, results for both these approaches.  For all computations in this work, we use Gaussian smearing because of its exponentially faster decay compared to Lorentzian smearing, which has long tails that slowly vanish at a rate proportional only to $1/x^{2}$, leading to poorer numerical stability. Now, by choosing a particular FWHM for Gaussian \eqref{eq:delta-g} and a given BZ grid of wavevectors for the summations over $\mathbf{q}$ and branches $j$ we can construct all superoperator matrices entering \eqref{eq:rr-super-op} and solve it as a system of linear equations for components of $\Xi_\mathrm{R}(\omega)$, i.e.
\begin{equation}\label{eq:lin-eq-xi}
    \left[\frac{i}{\mathrm{\hbar}}M_\mathrm{L} + \frac{1}{\mathrm{\hbar}^2}\hat{M}_{\mathcal{K}_\mathrm{r}}(0) - i\omega \mathrm{I}\right]\Xi_\mathrm{R}(\omega) = M_{\rho_\mathrm{A}(0)} + \frac{i}{\mathrm{\hbar}}\hat{M}_\mathrm{\psi}(0).
\end{equation}
Because of the spin-phonon coupling operator symmetry (S4.7.14), we only need to use BZ grid covering half of the reciprocal unit cell. Equally spaced grids within the whole BZ are extremely costly to properly represent acoustic and low-optical regions with considerable dispersion around the $\Gamma$ point. To tackle this problem, we propose to use an anisotropic onion-like multigrid, which was purposefully developed and described in Section S7.1 and schematically visualised in Figure S11. Here, we follow the convergence procedure proposed in \cite{Lunghi2023} for the relaxation time estimation, adapted to our susceptibility framework.
\begin{enumerate}[label=(\roman*)]
    \item We start with a large FWHM = 50 cm$^{-1}$ and compute susceptibility in d.c. static field of a given direction and strength, in temperature $T$ for $\mathrm{\Gamma}$ point only ($n_\mathrm{ref}$ = 1) by finding $\Xi_R(\omega)$ for a dense grid of magnetic field frequencies and using it in \eqref{eq:final-response-matrix-form}, where $X$ is an operator of dipole magnetic moment ($\mu = -\mathrm{\mu_B}(\mathrm{g_e}\hat{S}+\hat{L})$) in the direction of the d.c. field.
    
    \item From the simulated $\chi''(\omega)$ curve, we extract the frequency $\omega_{\max}$ at the maximum of the imaginary component and use it as an approximation to the inverse relaxation time $1/\tau$ (see Figure \ref{fig:0} for justification).
    
    \item Keeping the chosen FWHM fixed, we increase the density of the Brillouin-zone grid $n_{\mathrm{ref}}$ until $\log(\omega_{\max})$ in Hz remains within a window of $\pm 0.01$ for five consecutive refinements.
    
    \item Once convergence with respect to $n_{\mathrm{ref}}$ is reached for a given FWHM, we decrease the broadening (here by a factor of 0.6) and repeat the $n_{\mathrm{ref}}$ convergence step (iii).
    
    \item We continue this alternating refinement of FWHM and $n_{\mathrm{ref}}$ until the change in $\log(\omega_{\max})$ between successive converged FWHM values is below 0.01 for five consecutive refinements.
    
    \item After convergence of the relaxation time is reached, we additionally verify convergence of the numerical susceptibility values themselves.

\end{enumerate}
We note here that to speed up the procedure we can neglect $\hat{M}_\mathrm{\psi}$ term and higher-order initial correlation corrections to $M_{\rho_\mathrm{A}(0)}$ using only $(M_{\rho_\mathrm{A}(0)})_{ab,cd} = A_{a,c}\delta_{d,b} - \delta_{a,c}A_{d,b}$, as in the approximate method discussed in the Results section, since we are only interested in probing convergence with respect to phonon grid. Also, as here we only probe approximate relaxation times, later we fit the simulated susceptibility curves using the external program relACs \cite{Liberka2022}, developed earlier by some of the present authors, and employing the Havriliak–Negami model \eqref{eq:hn-model} to treat simulated data on equal footing as the experimental ones. The outcome of the above-described convergence procedure for the most demanding case of \textbf{2} characterized by the largest energy splitting covering the whole frequency range and at high temperature $T$ = 45 K for the grid denoted with $q_{\mathrm{ranges}}$ = [0.001,0.005,0.025,0.5] is presented in Figure S12. This particular grid type was chosen based on the dispersion and convergence rate analyses where a denser grid around $\mathrm{\Gamma}$ point starting at 1/20 of half reciprocal axis length (0.025 - fractional coordinate) is leading to a fast convergence and covers most of acoustic branch dispersion, while we also divided it further by 5 (0.005 and 0.001 points) until no differences in susceptibility are recorded for each $n_\mathrm{ref}$. The converged values of FWHM = 0.00304 cm$^{-1}$ and $n_\mathrm{ref}$ = 118 visible in Figure S12 are used for all following simulations in this manuscript. For numerical reasons, we must also restrict the phonon frequencies taken into account when evaluating \eqref{eq:m-rho-a}, especially when dealing with scalar functions \eqref{eq:int}. As the analytic forms from (S4.7.29) show they are unbounded with respect to the $\omega$ arguments and thus can grow indefinitely, the bigger the arguments are. This is caused because of the lack of quickly decaying Dirac-delta-like terms compared to \eqref{eq:j} and \eqref{eq:j-corr} in \eqref{eq:mk-hat} and \eqref{eq:mpsi-hat}. This problem stems from the fact that they originate from imaginary time Dyson expansion (see Appendix F) truncated at the second order, and this is completely analogous to expanding the decaying $e^{-x}$ function as a Taylor series up to quadratic terms, where the leading $x^2$ is growing to infinity as $x \to \infty$. Thus, as this term should represent only a small perturbative correction, it is valid only near the expansion point. Therefore, to obtain numerical stability we chose to bound the window of modes taken into account to FWHM used in the smearing function and neglect $\omega_{\mathbf{q},j}$ in \eqref{eq:m-rho-a} that do not satisfy $|\omega_{ab}\pm\omega_{\mathbf{q},j}|< 1000 \cdot \mathrm{FWHM}$, i.e., $\approx 3$ cm$^{-1}$ for our converged grid. This constitutes an explicit numerical validity window for this perturbative correction tied to a convergence procedure of the broadening FWHM and phonon grid density. In future works, it is desirable to carry out the expansion to higher orders or to search for more robust alternatives for approximating correlated $M_{\rho_A(0)}$. All the above-mentioned computations, starting from the automated scripts for displacement generations, running CP2K, and ORCA quantum-chemical software, parsing their outputs, forming Hessian, Dynamical matrices, wavefunction overlaps, Hamiltonian gradients, generating BZ multigrids, solving susceptibility equations, and performing additional analyses e.g., phonon Dispersion, weighted phonon DOS, IR spectra, plotting etc., were all implemented and are available in the development branch of the previously published Python package SlothPy, which also contains a variety of simulation routines of static magnetic properties discussed elsewhere. \cite{Zychowicz2024} 

{\bf Modeling of magnetic relaxation times.}
To extract relaxation times from experimental or ab initio simulated complex a.c. susceptibility, we use the Havriliak–Negami model \eqref{eq:hn-model} as implemented in the relACs software \cite{Liberka2022} which allows simultaneous fitting of the $\chi'(\omega)$ and $\chi''(\omega)$ curves presented in Figure S61. For benchmarking and reference, we also provide results for previously developed methods of simulating relaxation times using second- and fourth-order time-dependent perturbation theories within Markovian-secular quantum master equations for linear vibronic coupling as developed in \cite{Lunghi2022} and \cite{Lunghi2026}, mimicking the most recent implementation in the MolForge software, with the omission of the crystal-field parameter (CFPs) conversion step and adapted to our custom spin-phonon coupling operators evaluation based on \cite{Mariano2023} with our new BZ multigrid integration machinery. Notably, what we later call 1-phonon Secular-Markov relaxation times are calculated by constructing a second-order quantum master matrix as described in \cite{Lunghi2023}
\begin{equation}\label{eq:one-ph-sec-markow-red}
    \begin{aligned}     
    R^\mathrm{1-ph}_{ab,cd} &= -\frac{1}{\mathrm{\hbar}^{2}} \sum_{\mathbf{q},j} \Bigg\{ \sum_{e}\hat{G}^{\mathbf{q},j}(\omega_{ec})\delta_{bd} Y^{\mathbf{q},j}_{a,e}Y^{-\mathbf{q},j}_{e,c}\\& - \hat{G}^{\mathbf{q},j}(\omega_{ac})Y^{-\mathbf{q},j}_{a,c}Y^{\mathbf{q},j}_{d,b} - \hat{G}^{\mathbf{q},j}(\omega_{bd})Y^{-\mathbf{q},j}_{a,c}Y^{\mathbf{q},j}_{d,b}\\&+ \sum_{e}\hat{G}^{\mathbf{q},j}(\omega_{ed})\delta_{ca} Y^{\mathbf{q},j}_{d,e}Y^{-\mathbf{q},j}_{e,b} \Bigg\},
    \end{aligned}
\end{equation}
where
\begin{equation}\label{eq:g-one-ph}
    \begin{aligned}
    \hat{G}^{\mathbf{q},j}(\omega_{ab}) &= \frac{\mathrm{\hbar}\mathrm{\pi}}{\omega_{\mathbf{q},j}} \Big[\bar{n}_{\mathbf{q},j}\hat{\delta}_G(\omega_{ab}-\omega_{\mathbf{q},j}) \\& + (\bar{n}_{\mathbf{q},j}+1)\hat{\delta}_G(\omega_{ab}+\omega_{\mathbf{q},j})\Big] 
    \end{aligned}
\end{equation}
($\hat{\delta}_G$ is defined in \eqref{eq:delta-g}) and, to obtain the secular approximation, we set terms that do not satisfy $|\omega_{ac} + \omega_{db}| < \epsilon$ in some small numerical degeneration energy window $\epsilon$ to zero. Note that, as mentioned above, we treat the Gaussian smearing here in a fully symmetrical manner. The characteristic relaxation time $\tau$ is approximated as a negative inverse of the smallest (in amplitude) non-zero eigenvalue of matrix \eqref{eq:one-ph-sec-markow-red}. Finally, the relaxation times $\tau(T,H)$ extracted from experiment or ab initio simulations are treated as values of a multivariable function of temperature and magnetic d.c. field within the following phenomenological model \cite{Zychowicz2024}
\begin{equation}\label{eq:tau-model}
    \begin{aligned}
    &\tau^{-1}(T,H) = A_{\mathrm{dir}}TH^{m} + \frac{B_{1}\!\left(1+B_{3}H^{2}\right)}{1+B_{2}H^{2}}\\& + \tau_{0}^{-1} \exp \left(-\frac{U_{\mathrm{eff}}}{\mathrm{k}_{\mathrm{B}}T}\right) + D \frac{\exp\left(\frac{\mathrm{\hbar}\omega}{\mathrm{k}_{\mathrm{B}}T}\right)} {\left[\exp \left(\frac{\mathrm{\hbar}\omega}{\mathrm{k}_{\mathrm{B}}T}\right)-1\right]^{2}},
    \end{aligned}
\end{equation}
implemented in relACs software \cite{Liberka2022} as presented in Figure S62. In this equation, the first term corresponds to the direct one-phonon process that causes relaxation by a one-step transition between states of opposite magnetization whose energy gap is controlled by the applied d.c. field. \cite{AbragamBleaney1970,Standley2013} The second one is in the form of the Brons-van Vleck formula and represents a temperature-independent two-state Landau–Zener-type process mainly driven by internal hyperfine and dipolar oscillating magnetic fields known as the quantum tunneling of magnetization (QTM), where B1 is the zero-field tunneling rate, B2 corresponds to the suppression of the tunneling due to the bringing states out of the resonance window by the external magnetic field, and B3 represents the concentration of the sources of magnetic moments causing the transition. \cite{VanVleck1940,Gatteschi2003,Gatteschi2006,Tesi2016_2,Amjad2017} The third term is the well-known Orbach thermal activation process, which usually describes a cascade of phonon-assisted transitions constituting an effective energy barrier $U_{\mathrm{eff}}$. \cite{Orbach1961} The last term named a local-mode process (LMP) is a second-order Raman-type process, which occurs mainly for systems with energy splitting high enough so that at a given temperature, there is no population of phonons that could reach the first excited state. Then relaxation takes place by a simultaneous absorption and emission of degenerated (or nearly degenerated for the case of systems in a finite weak magnetic field) low-lying phonons with energies around $\mathrm{\hbar}\omega$, which approximately gives the temperature dependence in the form of the Fourier transform of the two-phonon correlation function. \cite{Briganti2021,Zychowicz2024,Eaton2001,Aravena2020,Amoza2021,Klemens1962} It is characteristic for this process to take a thermally activated Arrhenius form $D \exp \left(-\frac{\mathrm{\hbar}\omega}{\mathrm{k}_{\mathrm{B}}T}\right)$ for $\mathrm{k}_{\mathrm{B}}T << \mathrm{\hbar}\omega$, making it behave exactly as the Orbach process.

\section*{Results and discussion}

\begin{figure*}[htbp!]
  \centering
  \includegraphics[width=0.91\textwidth]{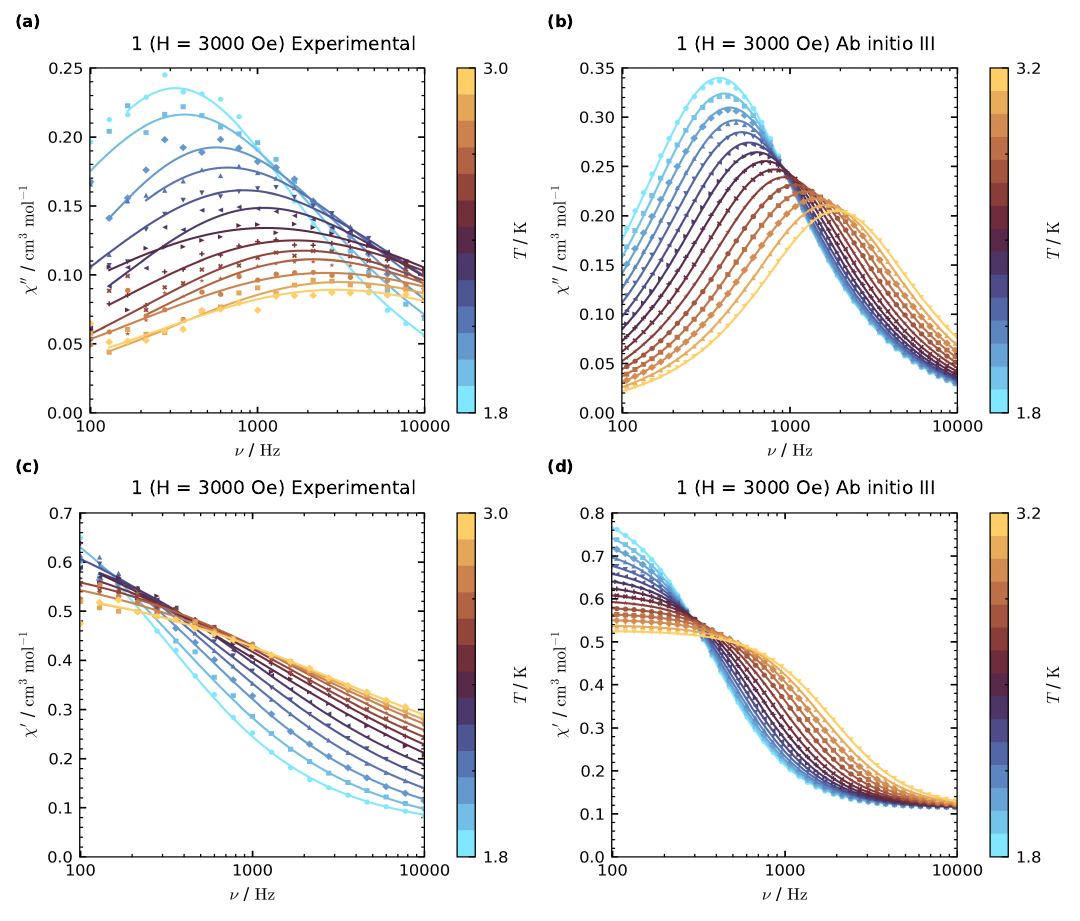}
  \caption{Temperature-variable experimental (a,c) magnetic characteristics of \textbf{1} under H$_{\mathrm{d.c}}$ = 3000 Oe compared to the corresponding Ab initio III simulation (b,d) averaged along $X$, $Y$, and Z magnetic axes for temperatures in the 1.8–3.2 K range, with experimental and ab initio data as points and solid lines representing the best-fit curves of the Havriliak–Negami model from \eqref{eq:hn-model}.}
  \label{fig:4}
\end{figure*}

\begin{figure}[htbp!]
  \centering
  \includegraphics[scale=0.91]{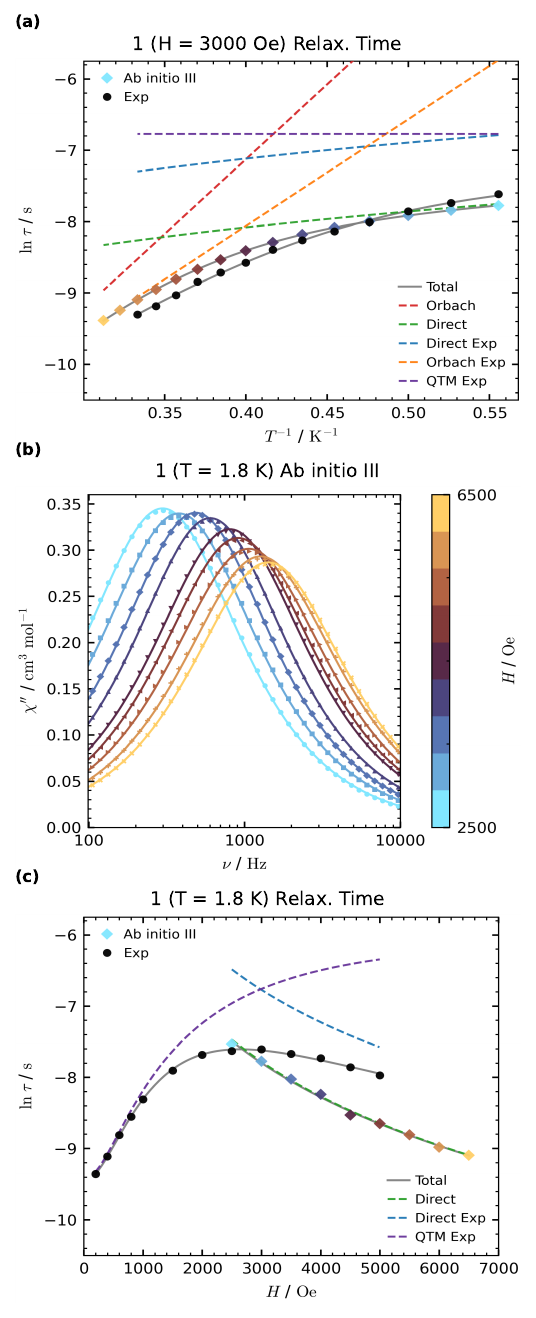}
  \caption{Temperature-variable H$_{\mathrm{d.c}}$ = 3000 Oe (a) and field-variable $T$ = 1.8 K (c) Ab initio III averaged over $X$, $Y$, $Z$ directions magnetic relaxation times $\tau$ of \textbf{1} compared to the experimental data (black circles) and fitted with model \eqref{eq:tau-model} where the colored dashed lines indicate the contribution of each relaxation path employed (the best-fit parameters are gathered in Table S9)  with corresponding field-variable imaginary component $\chi''$ (b) of the simulated complex a.c. susceptibility as points with lines representing the best-fit curves of the Havriliak–Negami model from \eqref{eq:hn-model}.}
  \label{fig:5}
\end{figure}

{\bf Ab initio a.c. magnetic susceptibility models and results for compound 1.} We begin our presentation with coordination polymer \textbf{1}, which experimental magnetic characterization is presented in Figures S13-S15 (see Section S6). The magnetically diluted lattice \textbf{1} does not show any out-of-phase imaginary a.c. susceptibility $\chi''(\omega)$ signal in the investigated frequency range in zero d.c. magnetic field. Only after the application of the bias magnetic field at a low temperature of 1.8 K (see Figure S14), the signal appears, and the relaxation time increases, which we interpret as the quenching of strong QTM caused by the highly mixed (non-axial) nature of the ground state. The relaxation time increases with the magnetic field, reaching its maximum around 3000 Oe, where the direct process between the split components of the ground-state Kramers doublet starts to dominate. For this value of the field, we conducted temperature studies presented in Figure S13, postulating an Orbach process as a dominant relaxation mechanism at higher temperatures (see also panels (a) and (c) of Figure \ref{fig:4} and panel (a) of Figure \ref{fig:5}). Figure S15 contains a three-dimensional visualization of the surfaces of processes from \eqref{eq:tau-model} as a simultaneous function of temperature and magnetic field with experimentally extracted relaxation times $\tau$.

\begin{figure*}[htbp!]
  \centering
  \includegraphics[width=0.91\textwidth]{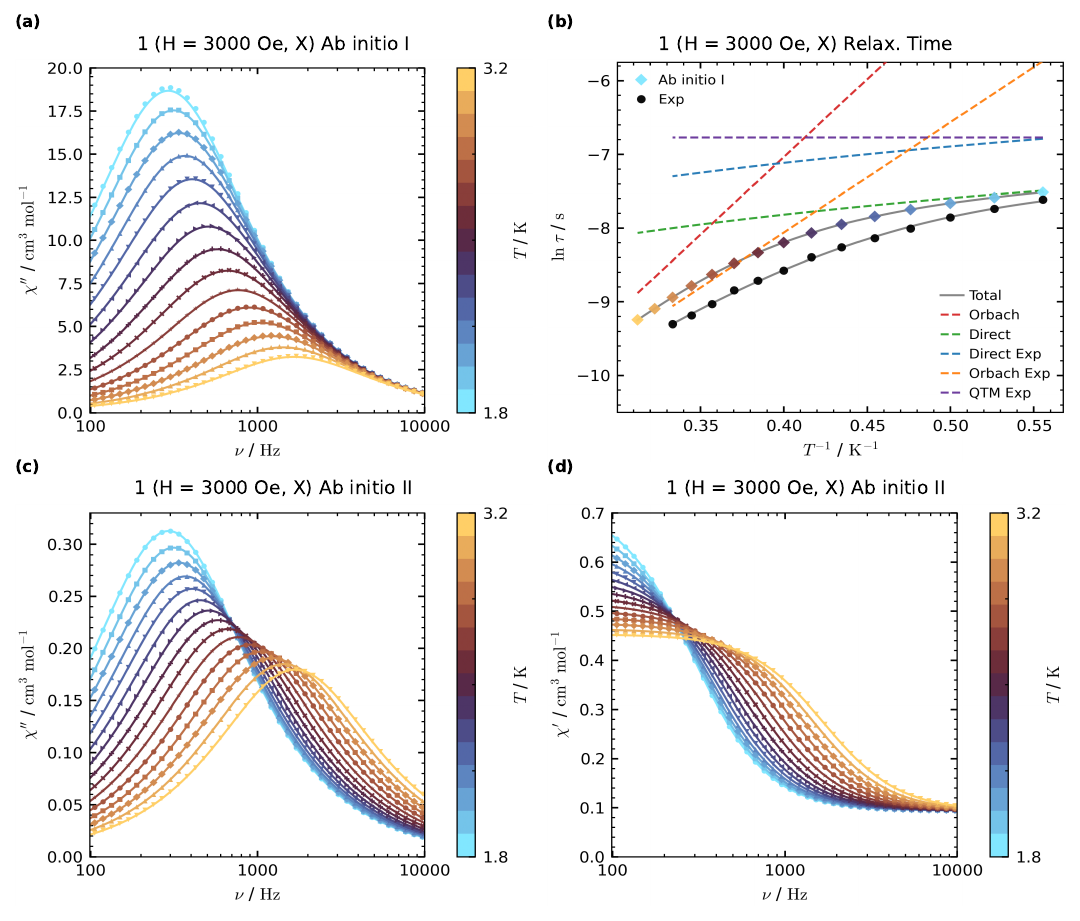}
  \caption{Temperature-variable Ab initio I (a) and II (c,d) a.c. magnetic characteristics of \textbf{1} under H$_{\mathrm{d.c}}$ = 3000 Oe applied along the main magnetic X-axis for temperatures in the 1.8–3.2 K range, with ab initio data as points and solid lines representing the best-fit curves of the Havriliak–Negami model from \eqref{eq:hn-model}, with the obtained magnetic relaxation times $\tau$ (b) and the fitted model \eqref{eq:tau-model} compared to the experimental data (black circles), where the colored dashed lines indicate the contribution of each relaxation path employed. The best-fit parameters are gathered in Table S9.}
  \label{fig:2}
\end{figure*}

\begin{figure*}[htbp!]
  \centering
  \includegraphics[width=0.91\textwidth]{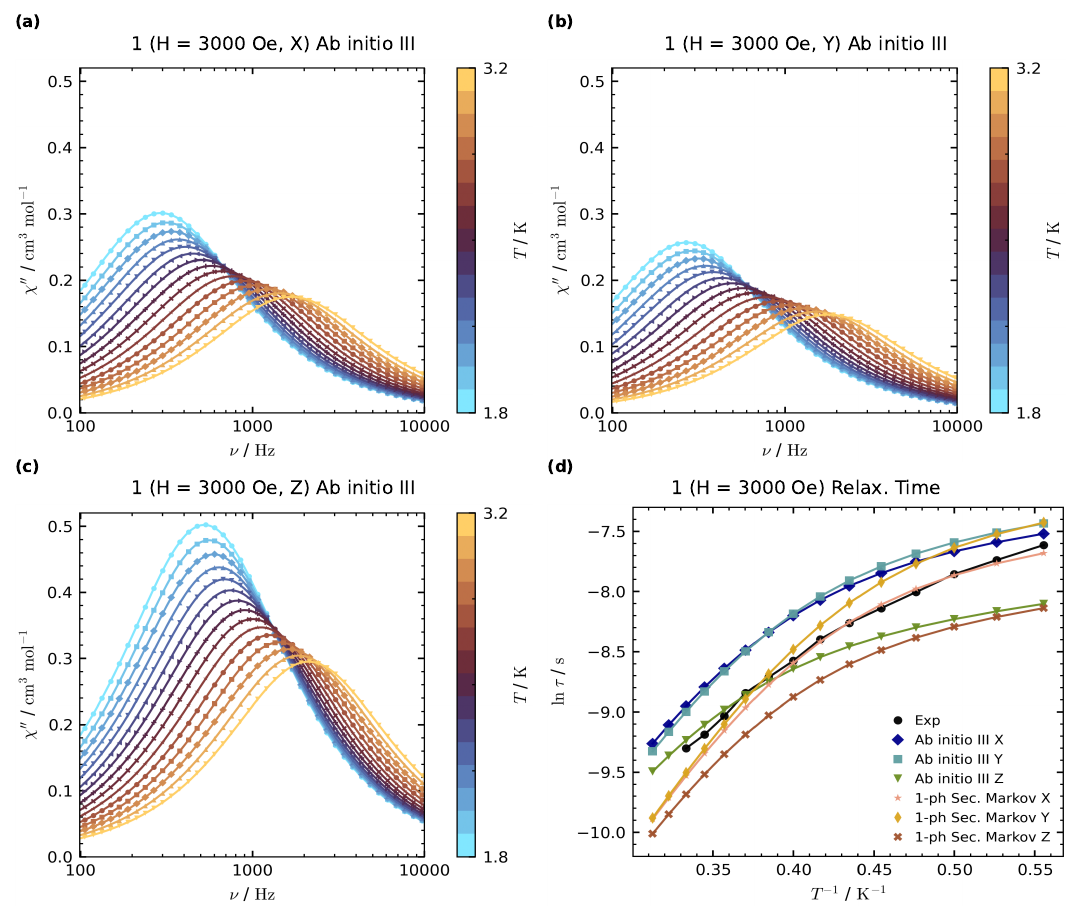}
  \caption{Temperature-variable Ab initio III (a-c) magnetic characteristics of \textbf{1} under H$_{\mathrm{d.c}}$ = 3000 Oe applied along the main magnetic $X$, $Y$, and $Z$ axes for temperatures in the 1.8–3.2 K range, with ab initio data as points and solid lines representing the best-fit curves of the Havriliak–Negami model from \eqref{eq:hn-model}, together with the obtained magnetic relaxation times $\tau$ compared to the experimental and those computed with one-phonon Secular-Markov approximation \eqref{eq:one-ph-sec-markow-red} presented in panel (d).}
  \label{fig:3}
\end{figure*}

To support our experimental interpretation and explore the role of each superoperator in \eqref{eq:rr-super-op}, we start our complex susceptibility simulations employing the computationally cheapest approximation, i.e., we set the matrix corresponding to the inhomogeneous term $\hat{M}_\mathrm{\psi}(0)$ in \eqref{eq:lin-eq-xi} to zero and take the uncorrelated limit for $M_{\rho_\mathrm{A}(0)}$, which results in
\begin{equation}\label{eq:ab-initio-I}
    \left[\frac{i}{\mathrm{\hbar}}M_\mathrm{L} + \frac{1}{\mathrm{\hbar}^2}\hat{M}_{\mathcal{K}_\mathrm{r}}(0) - i\omega \mathrm{I}\right]\Xi_\mathrm{R}(\omega) = M_{\rho_\mathrm{A}(0)},
\end{equation}
where $M_{\rho_\mathrm{A}(0)}\approx A\otimes I - I\otimes A^\mathrm{T}$ and $A = \mu_{\alpha}$ is the magnetic dipole moment operator in the direction $\alpha$ of the applied magnetic field. We further refer to this approximation as \textit{Ab initio I}. After obtaining the results of the simulations for a given temperature, frequency, and magnetic field ranges, we treat the data in the same manner as experimental ones by fitting with models \eqref{eq:hn-model} and \eqref{eq:tau-model} to extract relaxation times and phenomenological relaxation paths. The results of such an analysis for variable temperature are presented in Figures \ref{fig:2} - panels (a-b) and S16 for \textbf{1} under a static magnetic field $H_{\mathrm{d.c}} = 3000$ Oe applied in the direction of the main magnetic $X$-axis from Figure S10. 
Although the curves are perfectly described by a single-relaxation-time Debye model, i.e., \eqref{eq:hn-model} with $\beta = 1$ and $\alpha = 0$, the real susceptibility $\chi'(\omega)$ values are orders of magnitude too large and do not exhibit the expected limits of the isothermal $\chi_T$ and adiabatic $\chi_S$ susceptibilities for low and high frequency, respectively. The same is visible for $\chi''$ values that reveal an unexpected temperature dependence that causes the curves for each temperature not to cross, i.e., each line stays in the envelope of the previous (lower temperature) signal, forcing them to decay very rapidly. This behavior is attributed to the inconsistent truncation with respect to the perturbative order. Looking at the exact definition (S4.1.12) of the inhomogeneous term $\psi(t)$ entering NZ equation \eqref{eq:rrdo-diff-eq}, we see that it is non-zero only if $\mathcal{Q}[A,\rho_\mathrm{eq}] \neq 0$. Setting it to zero is equivalent to taking a factorized Born state as equilibrium $\rho_\mathrm{eq} \approx \rho_\mathrm{S}^\mathrm{eq} \otimes\rho_\mathrm{B}^\mathrm{eq}$ (see (S4.5.7)), completely disregarding $H_I$ (zeroth-order approximation). At the same time, the entire theory and therefore the propagator $U_0^{\mathcal{Q}}(t,s)$ entering definition (S4.1.13) of the response memory kernel $\mathcal{K}_\mathrm{r}(t-s)$ are formulated with respect to Hamiltonian \eqref{eq:h-0} containing spin-phonon interaction term $H_\mathrm{I}$. Because we approximate the kernel up to the second order in interaction $H_\mathrm{I}$, the same is true for the resolvent in \eqref{eq:rhoA-operator-form}, or equivalently in our static $\omega$ approximation \eqref{eq:lin-eq-xi}, which now acts on a source term $M_{\rho_\mathrm{A}(0)} + \frac{i}{\mathrm{\hbar}}\hat{M}_\mathrm{\psi}(0)$ approximated to an incorrect order giving completely wrong boundary conditions of static susceptibility and distorted temperature scaling. We see here that the initial correlations are crucial in the presented theory. However, since the curves perfectly follow model \eqref{eq:hn-model} we can easily restore the correct boundary behavior by first normalizing each signal $\chi(\omega, T)$ to the dimensionless unit transfer function $T(\omega,T)$
\begin{equation}
    T(\omega,T) = \frac{\chi(\omega, T) - \chi'(\omega \to \infty, T)}{\chi'(\omega \to 0, T) -\chi'(\omega\!\to\!\infty,T)},
\end{equation}
where we first subtract the adiabatic limit, obtained from the high-frequency limit of the ab initio simulated real susceptibility, and then divide by the low-frequency minus high-frequency limit to normalize $\chi'$ to unity. Even assuming $\beta \neq 1$ and $\alpha \neq 0$ in \eqref{eq:hn-model}, $T(\omega,T)$ should now have a form
\begin{equation}
    T(\omega,T) = \frac{1}{(1-(i\tau\omega)^{1-\alpha})^{\beta}}
\end{equation}
and we can simply restore the desired limits by performing the inverse operation
\begin{equation}\label{eq:ab-initio-II}
    \chi_{\text{Ab initio II}}(\omega,T) = T(\omega,T) \cdot (\chi_\mathrm{T}-\chi_\mathrm{S}) + \chi_\mathrm{S}.
\end{equation}
To computationally approximate $\chi_\mathrm{S}$ and $\chi_\mathrm{T}$ we first note that for $\omega \to \infty$ both $\hat{M}_{\mathcal{K}_\mathrm{r}}(\omega)$ from \eqref{eq:mk-hat} and $\hat{M}_\mathrm{\psi}(\omega)$ from \eqref{eq:mpsi-hat} vanish and we are only left with the unitary evolution part in \eqref{eq:rr-super-op}. Assuming a weak spin-phonon coupling we can set $\rho_\mathrm{eq} = \rho_\mathrm{S}^\mathrm{eq}\otimes\rho_\mathrm{B}^\mathrm{eq}$, i.e., $M_{\rho_\mathrm{A}(0)} = A\otimes \mathrm{I} - \mathrm{I}\otimes A^\mathrm{T}$ where $A = \mu_{\alpha}$ and according to Section S.4.5 we obtain an easy to-compute closed Kubo formula limit. This is the true adiabatic limit of our theory, where there is no energy dissipation and exchange with the lattice, and the evolution is purely unitary, which corresponds to an instantaneous response part in \eqref{eq:mag-adiabatic}, as opposed to slowly relaxing $m(t)$. The value of $\chi_\mathrm{S}$ is, therefore, decided by the commutator $[H_\mathrm{S}, \mu_{\alpha}]$, i.e. $\chi_S=0$ if the system's Hamiltonian commutes with the magnetic moment operator (it is diagonal in its basis). This is equivalent to the widely adopted requirement for the best performing SMMs \cite{Ungur2016,Vieru2024} where one desires not mixed, highly axial, energy states described by a single pseudo-m$_\mathrm{J}$ component in the magnetic moment basis, blocking any potential magnetic transitions between them. We show later whether this is true for our SMMs. The isothermal limit $\chi_\mathrm{T}$ is estimated within this approximation by calculating the magnetization expectation value with respect to the system's Hamiltonian $H_\mathrm{S}$, i.e. $M(T) = \mathrm{Tr}_\mathrm{S}(\mu_{\alpha}\rho_\mathrm{S}^\mathrm{eq})$, and divide it by the magnetic field in the finite differences approximation for static susceptibility valid for small fields or use a stencil differentiation as discussed in \cite{Zychowicz2024} and implemented in our SlothPy software. We label the susceptibility curves obtained that way from \eqref{eq:ab-initio-II} as \textit{Ab initio II}. In other words, the \textit{Ab initio I} model is used to determine the (dimensionless) transfer function $T(\omega,T)$ and thus the characteristic time scale, while the \textit{Ab initio II} model enforces independently computed static limits $\chi_\mathrm{T}$ and $\chi_\mathrm{S}$. Finally, we call \textit{Ab initio III} the model where the full second-order theory with the initially correlated states presented in the Theory section is used. The results of the \textit{Ab initio II} model for \textbf{1} under the magnetic field in $X$-axis direction are presented in panels (c-d) of Figure \ref{fig:2} and Figure S17. We note here that the described procedure only scales $\chi''$ values, leaving the frequency-dependent position of maxima unchanged; therefore, it has no effect on the extracted relaxation times $\tau$. The results of restoring all the terms in \eqref{eq:lin-eq-xi} up to the second order in spin-phonon interaction (\textit{Ab initio III}) for \textbf{1} in the $X$-direction of the field are presented in Figure S18 and panel (a) of Figure \ref{fig:3}.
We see that the expected frequency limits and, thus, temperature dependence from the \textit{Ab initio II} model are almost perfectly restored, which confirms the correctness of our implementation and shows that the spin-phonon coupling is indeed weak (as frequently approximated) in the sense of modifications to the initially correlated state, as it is not that much different from the limits provided for the factorized state. In addition, we do not observe any noticeable (in the limit of our fitting uncertainties) changes in the extracted relaxation times. Since the positions of maxima on the $\chi''$ plots are determined in our theory by the poles (characteristic relaxation rates in the linear-response sense) dictated by the resolvent from \eqref{eq:lin-eq-xi} this means that we deal with a single dominant pole or a combination of very close lying ones so that the modification to the source vector part provided by $\frac{i}{\mathrm{\hbar}}\hat{M}_\mathrm{\psi}(0)$ and corrections to $\rho_A(0)$ scale them uniformly acting as a multiplicative normalization factor and a value shift of the real component $\chi'$. This is analogous to the established ab initio theory of relaxation, where \eqref{eq:one-ph-sec-markow-red} is precisely an approximation to our response memory kernel $\mathcal{K}_\mathrm{r}$, providing that the dissipative, non-unitary, slow relaxation part of the equations can be approximated to a single relaxation time, usually the closest to null (equilibrium) eigenvalue corresponding to the slowest occurring relaxation process. \cite{Lunghi2023} However, the approach described here is more general and \eqref{eq:one-ph-sec-markow-red} is formulated as an effective evolution operator for the whole reduced density matrix, while in our current approach the memory kernel $\mathcal{K}_\mathrm{r}$ acts on RRDO defined in \eqref{eq:rrdo-def} which encodes in its structure the specific perturbing input proportional to $A$ (in our case magnetic dipole moment operator $\mu$) and further, through \eqref{eq:open-response-rrdo}, a specific recorded output $X$ - also $\mu$. We completely lose this information by using a pure quantum master equation approach, and therefore, we do not have any information about the relative weights of each eigenvalue/pole that could be visible during a.c. magnetic measurements. We note that our linear response implementation is capable of predicting multiple separate maxima on $\chi''$ plots corresponding to different relaxation processes whose relative intensity is decided by the $\mu$ operator, which was observed during FWHM convergence procedures. However, in the discussed case, as mentioned above, one or a couple of very closely lying poles dominate the relaxation, giving a single signal consistent with the experimental characteristics. As described in \cite{Lunghi2023}, one can mitigate such limitations of quantum master equations by indirectly introducing a magnetic field by studying the evolution of the full magnetization. However, this is more suitable for recreating d.c. magnetization remanence experiments than a.c. measurements.

We now move to the investigation of a.c. susceptibility simulations with respect to different directions of the applied d.c. magnetic field. Because \textbf{1} is a rather weakly performing SMM, which only exhibits field induced slow relaxation in a narrow low-temperature range, we observe a large adiabatic limit of the $\chi'$ signal and a significant response for each $X$, $Y$, and $Z$ directions along main magnetic axes, which explains the experimentally observed strong QTM persisting over a broad magnetic-field window. Regarding the non-negligible response in all directions together with the fact that the experiments are carried out using a powder sample, and since the a.c. susceptibility is a second-rank tensor (exactly as the static isothermal susceptibility from d.c. measurements), we should average the ab initio signals over $X$, $Y$, and $Z$ directions. We can do this because, thanks to the developed methodology, we know their exact absolute values (intensities). To accomplish this, we performed calculations within the \textit{Ab initio III} framework by applying the field along all three magnetic axes directions. Results are presented in panels (a-c) of Figure \ref{fig:3} and Figures S18-S20. At this point, since each of the calculations is performed for a single direction, we can compare the relaxation times extracted by fitting of \eqref{eq:hn-model} for $X$, $Y$, and Z with those obtained from quantum master equations. Such a summary is presented in panel (d) of Figure \ref{fig:3} and in the corresponding (d) panels of Figures S18-20 with the best-fit parameters of model \eqref{eq:tau-model} gathered in Tables S9 and S10. Both approaches reveal the same effective thermal activation barriers $U_{\mathrm{eff}}$ of the fitted Orbach processes that vary with the direction of the magnetic field from 14.6 to 14.9 cm$^{-1}$ that correspond to the changes of energy splitting between the ground and the first excited doublet (14.4 cm$^{-1}$; Table S8) caused by the Zeeman interaction with the field. However, they differ in the absolute values of the $\tau_{0}^{-1}$ and $A_{\mathrm{dir}}$ factors determining $\tau$ (the values of the field exponent $m$ are addressed in the subsequent discussion), as well as in their relative magnitudes that represent the strength between the direct and Orbach processes. As there are no clear trends to highlight here, we note that all the results are spread within one order of magnitude from the experimental relaxation time data, and their differences cannot be neglected, adding to the list of possible discrepancies of Secular-Markov approach commonly observed when compared to the a.c. magnetic data. \cite{Goodwin2017,Mariano2023,Briganti2021,Mondal2022,Kragskow2023,Lunghi2023} We see that the \textit{Ab initio III} relaxation times along each direction differ significantly, as depicted by the positions of the  $\chi''$ maxima in Figure \ref{fig:3} and, intensity-wise, are dominated by the $Z$ component giving the largest magnetic response (about 0.5 cm$^3$ mol$^{-1}$ for 1.8 K). The results of their averaging are presented in Figure \ref{fig:4} along with the experimental data and in more detail in Figure S21.
Even after adding curves when averaging over the directions of the applied field, the result is still perfectly described with a single-relaxation time \eqref{eq:hn-model} with only a slightly distorted parameter $\alpha$ with a maximal value along different temperatures $\alpha=0.02$. Visually, this is attributed to the fact that both $\chi''$ curves for the $X$ and $Y$-directions, for each temperature, are contained within the envelope of the strongest $Z$-axis signal, and no separate maxima are observed as postulated for closely-lying resolvent poles before. Comparison with experiment reveals satisfactory agreement of the isothermal limit $\chi_\mathrm{T}$ while the experimental curves are characterized by large $\alpha$ and $\beta \neq 1$ parameters describing their extremely elongated oblate shape encoding a large distribution of relaxation times with a huge adiabatic limit not described within our theory and attributed to various defects of a real crystal lattice and experimental setup. Note that the value of $\chi''$ scales as $\chi_\mathrm{T} - \chi_\mathrm{S}$. However, the extracted relaxation times presented in panel (a) of Figure \ref{fig:5} and Figure S21 are exceptionally well in agreement with the experimental data with $U_{\mathrm{eff}}$ of 10.4 and 14.9 cm$^{-1}$ for experiment and ab initio simulation, respectively.
We once again note that such an averaging would not be possible if one were to only know the sole relaxation times for different directions without the numerical value of the complex a.c. susceptibility itself.

{\bf Field-variable ab initio results for compound 1.}
In order to obtain the full picture of the relaxation mechanisms and to obtain $\tau(T,H)$ as a multivariate function of temperature and magnetic field, we perform analogous \textit{Ab initio III} averaged studies, but in the function of the strength of the applied magnetic field at low 1.8 K temperature to match the experimental condition. The results of such simulations are presented in panels (b-c) of Figure \ref{fig:5} and Figure S22, while the parameters of the fitted model \eqref{eq:tau-model} for the relaxation times obtained by simultaneous analysis of temperature- and field-dependent relaxation times are gathered in Table S9 with a three-dimensional view of surfaces corresponding to different relaxation paths visualized in Figure S23. The field-variable computations reveal a monotonous decrease of relaxation times with increasing magnetic field, expected for a direct process due to the increasing energy gap between the states of the ground doublet connected with better phonon availability. The success of such a fully ab initio field-dependent direct process simulation originates from the developed BZ multigrid described in Section S7.1, which provides perfect coverage of the acoustic phonon region for the studied coordination polymer with phonon energies from 0.3 to 0.5 cm$^{-1}$ corresponding to the Zeeman field-induced energy gap between the ground doublet states. The fitted magnetic field exponent $m$ from model \eqref{eq:tau-model} has value 1.71(2) and is close to the experimentally found 1.57(8) (see Tables S9) and that predicted from a Debye phonon dispersion model for the acoustic region for Kramers systems $m = 2$ \cite{Lunghi2023}. We note here that this is a value that originates from the direction-averaged studies, and it varies for different magnetic axes. For reference we report the results for the main magnetic axis $X$ in Figure S24 together with the corresponding Secular-Markov result for a single direction where both dependencies follow a similar course and are characterized with $A_{\mathrm{dir}} = 2.34(1) \cdot10^{-3}; 2.77(1) \cdot10^{-3}$ and $m = 1.63(7); 1.61(4)$ for the \textit{Ab initio III} and quantum master equations, respectively. Nevertheless, for fitting only temperature-variable data containing the direct process as a component, we fix the value of $m$ to 1.71 because for the static field this acts only as a constant factor, which makes $A_{\mathrm{dir}}$ directly reflect differences in direct process strength along the sets of data (see Tables S9 and S10). One also needs to be careful when comparing the results to the experimental characteristic, as it is primarily dominated by a strong QTM component not present within our theoretical framework. The expected field dependence of the QTM is an increase in relaxation time with magnetic field, due to an increased energy gap between states that effectively brings them out of the resonance window. We do not observe that in our simulations, as it would require an inclusion of a bias field perpendicular to the applied a.c. field present in the crystal along with dipolar/hyperfine oscillating fields of frequencies comparable to the states' energy gaps and phonon frequencies in the interaction Hamiltonian. The only truly temperature-independent response pathway present in our current theory is described by the poles of the purely unitary resolvent from \eqref{eq:lin-eq-xi} $\left[\frac{i}{\mathrm{\hbar}}M_\mathrm{L} - i\omega \mathrm{I}\right]$ obtained after setting to zero $\hat{M}_{\mathcal{K}_\mathrm{r}}$ and $\hat{M}_\mathrm{\psi}$ phonon-dependent parts. Those are infinitely sharp peaks not broadened by the bath present at field frequencies corresponding to energy gaps of states which, as discussed earlier, in the applied magnetic field are way beyond the available applied a.c. magnetic field frequencies. For reference, we also provide the corresponding temperature-variable results of the \textit{Ab initio III} simulations under $H_{\mathrm{d.c}}$ = 3000 Oe in $X$, $Y$, and $Z$ axes directions in Figures S25-S27 and Tables S9, S10, but computed treating the broadening functions in a symmetric manner as discussed in the Computational Methods section (labeled as \textit{Ab initio III Symm.}) to ensure that the differences between relaxation times of our linear response theory and quantum master equations treatment are not solely caused by the change in the smearing procedure. Interestingly, relaxation times obtained in this way are consistently faster, by approximately an order of magnitude, than all of the discussed theoretical simulations and the experimental relaxation times, effectively providing the worst model.

Even though our ab initio simulations, based only on one-phonon processes, accurately describe the experimental results, one could expect that in such low temperatures regime, a two-phonon Raman-type relaxation would be dominant. To verify this, we performed the simulation of two-phonon contributions through fourth-order quantum master equations based on linear-vibronic coupling, with an implementation equivalent to that available in the MolForge package \cite{Lunghi2022} described in \cite{Lunghi2023} and recently \cite{Lunghi2026}, but customized to our BZ multigrid (Section S7.1) and spin-phonon operator machinery. As this type of calculations are extremely costly, rigorous convergence of the results with respect to FWHM and BZ grid is beyond the scope of this manuscript. To mitigate that and obtain an approximation, we chose a large enough FWHM = 10 cm$^{-1}$ so that we are able to convincingly converge the relaxation times for not too large values of $n_\mathrm{ref}$ for our nested multigrid with $q_{\mathrm{ranges}}$ = [0.001,0.005,0.025,0.5] covering the acoustic region. This procedure is presented in Figure S28, where for $n_\mathrm{ref}$ = 13 the stabilization is reached. Further, Figures S29-S31 contain a comparison of the resultant Raman relaxation times for $X$, $Y$, and $Z$ directions of the applied magnetic field H$_{\mathrm{d.c}}$ = 3000 Oe with experimental times and those obtained for all other discussed ab initio models. As the absolute values of the Raman relaxation times obtained for a single chosen FWHM are not that reliable, we note that the slope of the dependence, encoding energies of phonons taking part in the relaxation, versus inverse temperature, is way too large compared to the experimental one and other approaches, which supports our one-phonon relaxation picture.

\begin{figure}[htbp!]
  \centering
  \includegraphics[scale=0.91]{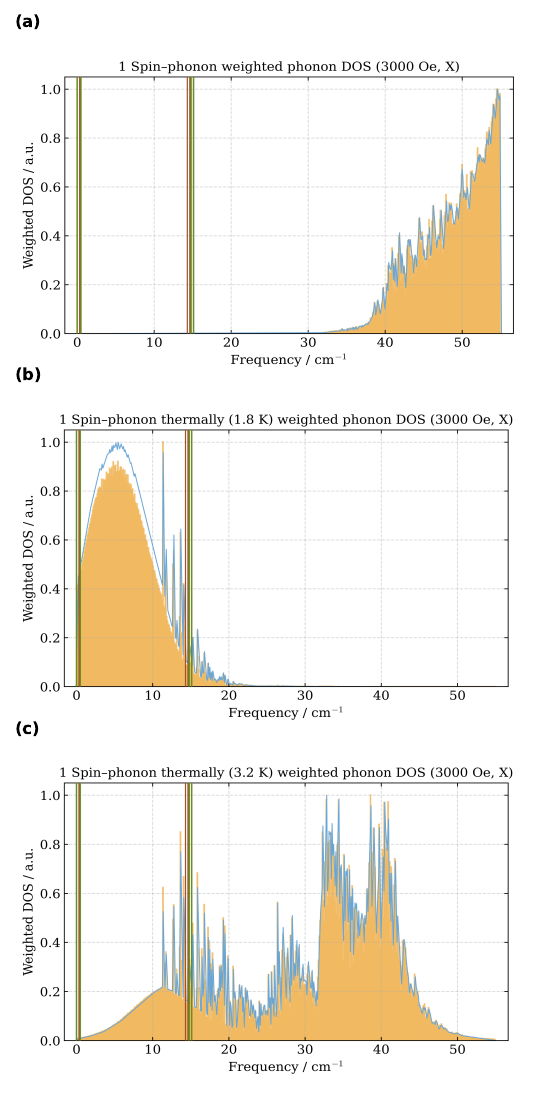}
  \caption{Spin-phonon weighted phonon density of states (DOS) of \textbf{1} under H$_{\mathrm{d.c}}$ = 3000 Oe applied along $X$ axis (a) with thermally weighted equivalent at 1.8 K (b) and 3.2 K (c) constructed according to Section S7.2 with energy levels split by the magnetic field in a given frequency window indicated as green lines and all possible energy transitions as red lines for BZ multigird with $q_{\mathrm{ranges}}$ = [0.001,0.005,0.025,0.5] and $n_\mathrm{ref}$ = 118 where the yellow part is a weighted histogram counting phonons and blue line represents a Gaussian smearing with FWHM = 0.1 cm$^{-1}$.}
  \label{fig:6}
\end{figure}

A valuable additional interpretational tool is provided by ordinary phonon density of states (DOS) plots, as well as spin-phonon and thermally weighted equivalents constructed for the used BZ multigrid as described in Section S7.2 in the frequency window corresponding to the energy of quantum states that take part in the transitions. Figure \ref{fig:6}, along with Figures S32-S35 contain such a summary.
It is then apparent that the relative density of phonons, even those weighted by the corresponding spin-phonon operator norm, for the relevant energy transitions is negligible compared to the optical neighborhood above 30 cm$^{-1}$, as those are completely within the acoustic region. However, this picture changes dramatically after thermal phonon occupation is accounted for. We see that at such a low temperature of 1.8 K, the direct transition within the first doublet with energy around 0.5 cm$^{-1}$ is preferred over the transition to the first excited one around 14.7 cm$^{-1}$. After increasing the temperature to 3.2 K, the direct relaxation pathway is no longer the strongest, as it is clearly dominated by the thermal activation Orbach-like transition to the first excited doublet as seen from the DOS changes in Figure \ref{fig:6} between panels (b) and (c). It is therefore consistent with the proposed phenomenological factorization of the temperature dependence of relaxation times on the direct and Orbach components. We can also note that the low-lying optical modes (sharp spikes above 10 cm$^{-1}$ in panels (b), (c) of Figure \ref{fig:6}) are characterized by large spin-phonon coupling compared to the acoustic continuum appearing at extremely low frequencies. Those pictures indicate the pivotal role of the developed custom BZ multirid in our simulations, as it provides effective coverage of the crucial acoustic regime that is active at low temperatures and is unattainable with ordinary uniform grids. The last numerical experiment presented in Figure S36 is to exclude phonons lying below 0.7 cm$^{-1}$ (the highest energy gap between two ground states in 3000 Oe field along $Z$ direction is 0.65 cm$^{-1}$) from our \textit{Ab initio III} simulations. Under these conditions, we see that the direct process is quenched and the relaxation time follows the previously fitted Orbach dependence over the whole temperature range, confirming our previous interpretation of the relaxation mechanism. 

\begin{figure*}[htbp!]
  \centering
  \includegraphics[width=0.91\textwidth]{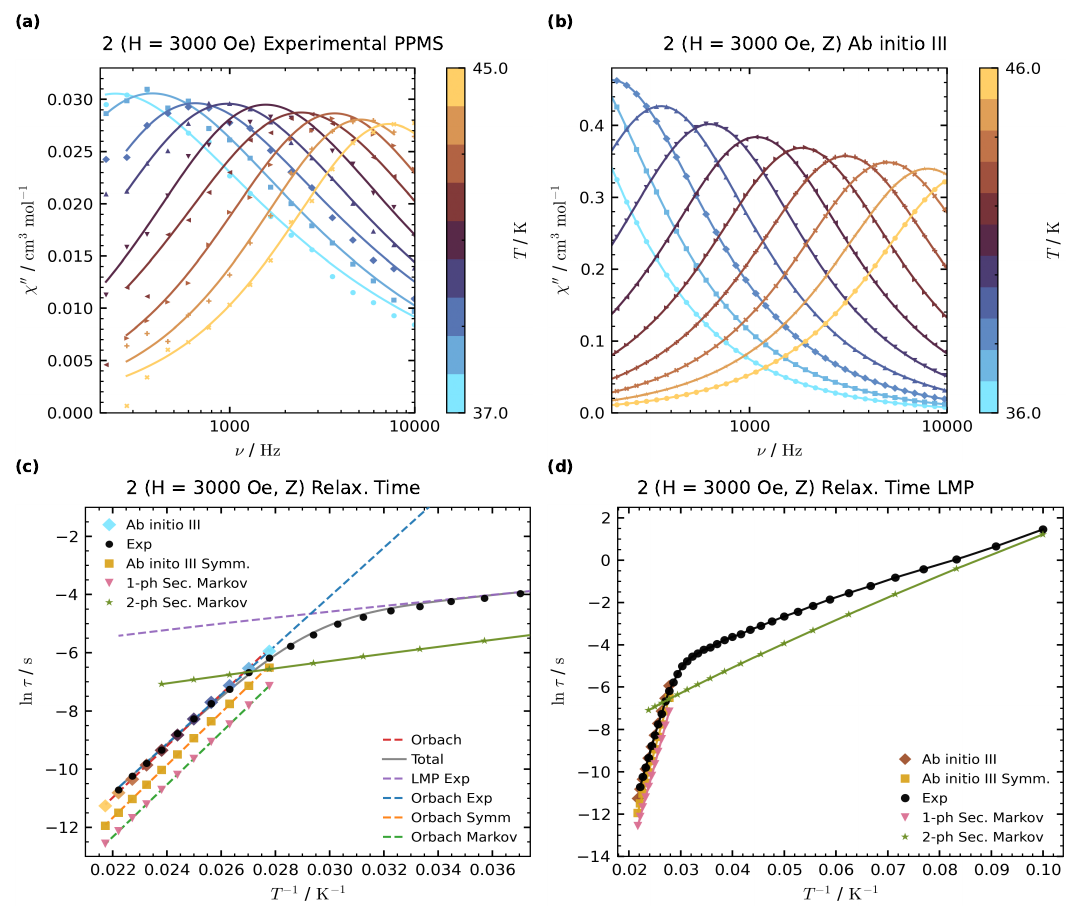}
  \caption{Temperature-variable experimental magnetic characteristics (a) of \textbf{2} under H$_{\mathrm{d.c}}$ = 3000 Oe recorded for the high-temperature region compared to the corresponding Ab initio III simulation (b) for temperatures in the 36–46 K range, with experimental and ab initio data as points and solid lines representing the best-fit curves of the Havriliak–Negami model from \eqref{eq:hn-model} together with magnetic relaxation times $\tau$ for various considered ab initio models (c) for high-temperature regime compared to the experimental data (black circles) and fitted with model \eqref{eq:tau-model} where the colored dashed lines indicate the contribution of each relaxation path employed and the overview covering the whole experimentally investigated temperature range (d) highlighting the Raman-type simulated LMP. The best-fit parameters are gathered in Table S11.}
  \label{fig:7}
\end{figure*}

{\bf Ab initio simulations and results for compound 2.} We also performed an analogous set of ab initio simulations for \textbf{2}, which was already structurally and magnetically characterized in \cite{Wang2023} under H$_{\mathrm{d.c}}$ = 3000 Oe and a.c. magnetic field in the frequency range of 0.1 to 1000 Hz as presented in Figure S37. We re-fitted reported experimental data using the full Havriliak–Negami model \eqref{eq:hn-model} and changed the original interpretation based on the double-Raman $T$-power law process reported in \cite{Wang2023} to a single LMP with $\mathrm{\hbar}\omega$ = 59.4(6) cm$^{-1}$ (Table S11) as suggested by the clear exponential $T$-dependence of magnetic relaxation time in the 10-30 K temperature range and large energy splitting of 295.1 cm$^{-1}$ between the ground and the first excited pseudo-doublets (Table S8). To gain additional insight into the high-temperature range above 36 K dominated by the Orbach process, we conducted magnetic experiments in the extended a.c. field frequency range of 100 to 10000 Hz. The discussed results are contained in panel (a) of Figure \ref{fig:7} and Figure S38, while the combined magnetic relaxation time plot is reported in Figure S39.
Once magnetically diluted, \textbf{2} becomes a very effective zero-dc field SMM for a broad temperature range. Coherently, \textbf{2} is magnetically very anisotropic and characterized by axial states giving negligible response in the $XY$ magnetic axis plane as compared to the $Z$ direction. We therefore report ab initio results only for the direction of the $Z$-main magnetic axis, as others give susceptibility a few orders of magnitude weaker and have virtually no effect within the averaging procedure. This is also accompanied by an adiabatic susceptibility approaching zero, as discussed earlier. Note that when comparing these single-axis calculations with powder measurements, the corresponding experimental powder-averaged susceptibility should be approximately one-third of the response of the $Z$-axis because the components $X$ and $Y$ are negligible on the present scale. The \textit{Ab initio I-III} results for the Orbach-dominated temperature range are presented in Figures S40-S42 where the extracted magnetic relaxation times are also compared to the experimental and Secular-Markov values, while we also report for reference \textit{Ab initio III Symm.} simulations with symmetric broadening in Figure S43, with best-fit parameters of model \eqref{eq:tau-model} for various approximations gathered in Table S11. Once again, we do not observe any quantifiable differences in relaxation times across the three models. They match the experimental characteristics very well, as seen in panels (a-c) of Figure \ref{fig:7}, while the symmetric and Secular-Markov Redfield-based approaches give three to four times faster relaxation times as quantified by the $\tau_{0}^{-1}$ parameter in Table S11. Interestingly, the effective energy barriers $U_{\mathrm{eff}}$ of the Orbach process differ slightly between various ab initio models, with 628(3) and 624.3(2) cm$^{-1}$ for symmetrical and secular models, and 617(5) cm$^{-1}$ for \textit{Ab initio III} being the closest to the experimental value of 585(8) cm$^{-1}$. This means that they numerically describe different compositions of the relaxation processes with \textit{Ab initio III}, to some extent, preferring the channel connected to the sixth energy state and the other through the seventh, from Table S8 whose energies are modified by the 3000 Oe magnetic field.

We note that it is difficult to compare numerical values of imaginary a.c. susceptibility $\chi''$ to experimental ones (panels (a-b) of Figure \ref{fig:7}) as those are characterized with experiment-dependent large adiabatic component which is close to zero within our theory and, as already mentioned, $\chi''$ scales as $\chi_\mathrm{T} - \chi_\mathrm{S}$. Furthermore, we notice that the \textit{Ab initio II} procedure gives an isothermal limit $\chi_\mathrm{T}$ consistent with the experimental data, while Ab initio III results in the expected limit only for temperatures above 39 K (Figures S38, S41, and S42). This is exactly the region where the one-phonon Orbach relaxation type dominates over the postulated LMP, with susceptibility values drifting away for lower temperatures, which is even more pronounced in the symmetric-type simulations that overall show a much poorer value convergence in Figure S43. This can be interpreted as an indicator that our second-order linear coupling equations are insufficient to approximate a temperature region presumably dominated by two-phonon Raman-like processes that need the expansion to be carried over to fourth-order in system-bath interactions $H_\mathrm{I}$. In other words, we might be out of the validity window of the second-order perturbative corrections, as seen from the simulated relaxation times deviating from the experiment at lower temperatures. The presented truncation inefficiency is a motivating factor to expand the presented theory up to the fourth leading order for the expansions of $\hat{M}_{\mathcal{K}_\mathrm{r}}$, $\hat{M}_\mathrm{\psi}(\omega)$, and $(M_{\rho_\mathrm{A}(0)}$ in future work.

\begin{figure}[htbp!]
  \centering
  \includegraphics[scale=0.91]{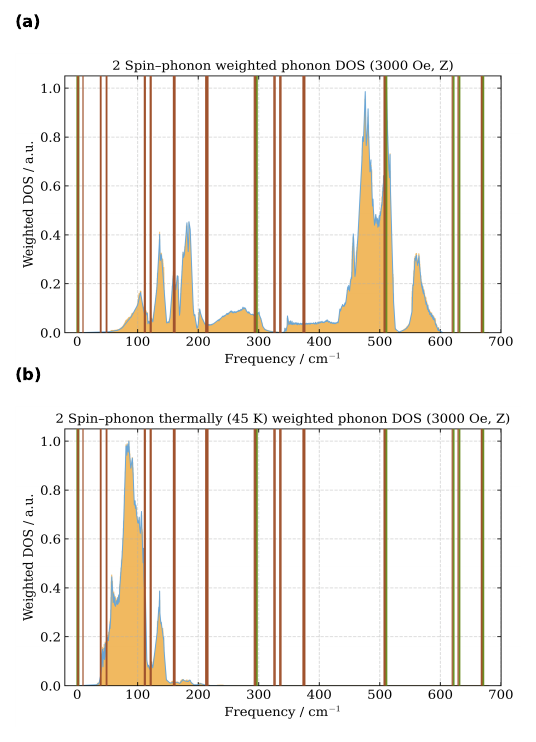}
  \caption{Spin-phonon weighted phonon density of states (DOS) of \textbf{2} under H$_{\mathrm{d.c}}$ = 3000 Oe applied along $Z$ axis (a) with thermally weighted equivalent at 45 K (b) constructed according to Section S7.2 with energy levels split by the magnetic field in a given frequency window indicated as green lines and all possible energy transitions as red lines for BZ multigird with $q_{\mathrm{ranges}}$ = [0.001,0.005,0.025,0.5] and $n_\mathrm{ref}$ = 118 where the yellow part is a weighted histogram counting phonons and blue line represents a Gaussian smearing with FWHM = 0.1 cm$^{-1}$.}
  \label{fig:8}
\end{figure}

To support our interpretation, we perform analogous Raman simulations as for \textbf{1} within, the only currently available fourth-order framework, namely the one of Secular-Markov quantum master equations. The same Gaussian smearing with FWHM = 10 cm$^{-1}$ is used here, and the convergence with respect to our custom grid is presented in Figure S44. The obtained Raman magnetic relaxation times are compared to the experiment and other approaches in panel (d) of Figure \ref{fig:7} and Figures S45, S46. The comparison reveals satisfactory agreement with the experiment with LMP model parameters $D$ and $\mathrm{\hbar}\omega$ (Table S11) equal to 1093(57), 9003(89) s$^{-1}$ and 59.4(6), 71.6(7) cm$^{-1}$ for experiment and two-phonon ab initio simulations, respectively, and crossover with Orbach simulated relaxation times around 37 K. Deeper elucidation of the postulated mechanisms is acquired by the help of ordinary and spin-phonon weighted phonon DOS similar to those discussed already for \textbf{1} but spanning much larger frequency window. From the phonon DOS contained in Figure S47 we see that there are no vibrations available above a frequency of 610 cm$^{-1}$, which automatically excludes the possibility of direct transitions to excited states above this threshold. The only possibility is the involvement of states with energies lying around the highly populated 500 and 300 cm$^{-1}$ regions, which is also consistent with the spin-phonon weighted DOS presented in panel (a) of Figure \ref{fig:8} and Figure S48.
The picture changes dramatically after accounting for the phonon occupation numbers in the thermally weighted DOS from panel (b) of Figure \ref{fig:8} and Figure S49, where we see that even at a temperature of 45 K, the effective coupling strength around 300 cm$^{-1}$ is extremely low. This observation, connected with the fact that the direct transition within the ground Ising pseudo-doublet is also suppressed both by the magnetic states' axiality and spin-phonon coupling in 0-3 cm$^{-1}$ corresponding to their energy gap in 3000 Oe field, explains why \textbf{2} is such an excellent SMM. As a population transition to the second and third excited states lying around 300 cm$^{-1}$ takes place, it can be followed by a cascade of one-phonon transitions to the next excited states, as suggested by the much larger coupling strength in the 40-200 cm$^{-1}$ window constituting a true "ladder-like" Orbach process with effective energy barriers, as reported in Table S11. This also gives a hint about the origins of the observed LMP process, as even two-phonon relaxation is preferred to involve pairs of nearly degenerate phonons from the highly active region around 70 cm$^{-1}$ ($\mathrm{\hbar}\omega$ = 71.6(7) cm$^{-1}$ for the simulated LMP) rather than phonons resonant with the first excited state around 295 cm$^{-1}$.

\begin{figure}[htbp!]
  \centering
  \includegraphics[scale=0.91]{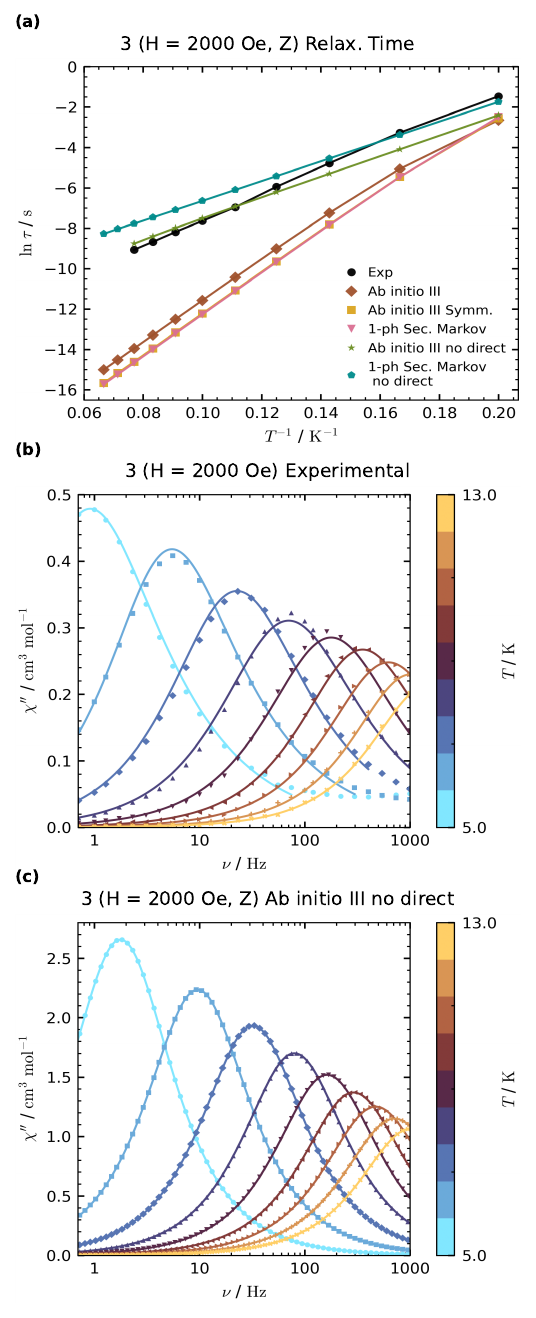}
  \caption{Temperature-variable experimental magnetic relaxation times of \textbf{3} (black circles) compared to various considered ab initio models (a) with experimental magnetic characteristics (b) under H$_{\mathrm{d.c}}$ = 2000 Oe compared to the corresponding Ab initio III simulation (c) obtained by neglecting phonon modes with frequencies above 69 cm$^{-1}$ for temperatures in the 5–13 K range, with experimental and ab initio data as points and solid lines representing the best-fit curves of the Havriliak–Negami model from \eqref{eq:hn-model}.}
  \label{fig:9}
\end{figure}

{\bf Ab initio simulations and results for compound 3.} Finally, we turn to an ab initio analysis of \textbf{3}, which was characterized in \cite{Xin2019}, and we report here additional experimental temperature-variable a.c. measurements for the magnetically diluted sample under a static d.c. field of 2000 Oe in Figure S50. The results of the Ab initio I-III Symm. procedures are presented in Figures S51-S54 with the best-fit parameters of the model \eqref{eq:tau-model} gathered in Table S12 where, due to the high anisotropy of \textbf{3} we once again report results only for the $Z$ main magnetic axis. As the summary of the magnetic relaxation times from panel (a) of Figure \ref{fig:9} shows, all of the ab initio models severely underestimate the experimental characteristic, resulting in much faster relaxation.
The data were interpreted using a thermally activated (Arrhenius-like) Orbach process from \eqref{eq:tau-model} giving $U_{\mathrm{eff}}$ for all models consistently around 71 cm$^{-1}$ compared to the experimental barrier of 43.7(6) cm$^{-1}$ and accompanied by $\tau_{0}^{-1}$ larger up to three orders of magnitude, as shown in Table S12. We note that we excluded from the fitting the last relaxation time point corresponding to a temperature of 5 K for all the models because it clearly does not follow the single exponential inverse temperature dependence. The extracted ab initio barriers suggest that the relaxation goes through the second excited doublet split by the magnetic field and not through the expected first one around 36.3 cm$^{-1}$ (see Table S8), which is closer in energy to the experimental barrier.

\begin{figure}[htbp!]
  \centering
  \includegraphics[scale=0.91]{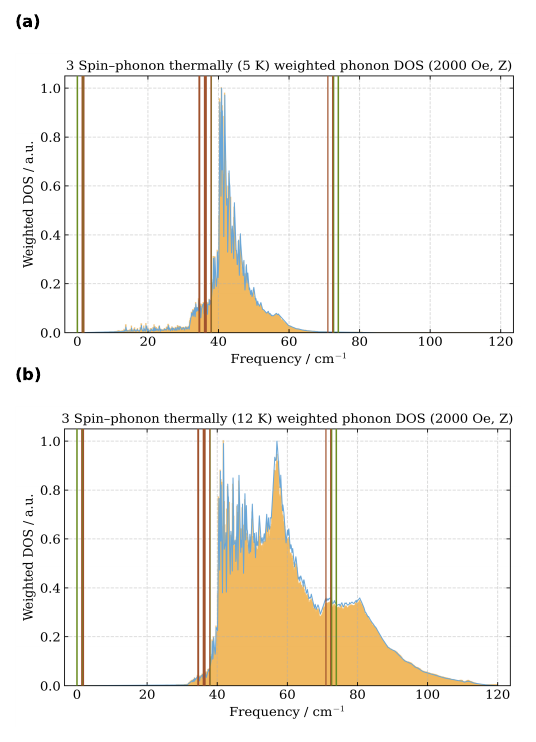}
  \caption{Spin-phonon thermally weighted phonon density of states (DOS) of \textbf{3} under H$_{\mathrm{d.c}}$ = 2000 Oe applied along $Z$ axis at 5 K (a) and 12 K (b) constructed according to Section S7.2 with energy levels split by the magnetic field in a given frequency window indicated as green lines and all possible energy transitions as red lines for BZ multigird with $q_{\mathrm{ranges}}$ = [0.001,0.005,0.025,0.5] and $n_\mathrm{ref}$ = 118 where the yellow part is a weighted histogram counting phonons and blue line represents a Gaussian smearing with FWHM = 0.1 cm$^{-1}$.}
  \label{fig:10}
\end{figure}

To reveal the nature of the observed ab initio relaxation, we conduct a numerical experiment which involves excluding from the simulation the lattice vibrations above the 69 cm$^{-1}$ frequency, i.e., slightly below the energy of the second excited doublet involved. The results of Ab initio III and the Secular-Markov procedure obtained in this way are contained in Figures S55, S56 and are compared to the experimental characteristics in panels (b-c) of Figure \ref{fig:9} with the best-fit parameters in Table S12. They give now an effective energy barrier $U_{\mathrm{eff}}$ around 35 cm$^{-1}$ corresponding to the first excited doublet in the magnetic field, which reveals that the extremely efficient ab initio relaxation process observed initially is, in fact, a direct one-phonon excitation to the state of the second excited doublet. This is because phonons below 69 cm$^{-1}$ are still available, and if the initial process consisted of multi-step Orbach-like transitions through lower states, it would still dominate the relaxation and give the same effective energy barrier. We also emphasize here how similar the simulation conducted within the limited frequency window is to the experimental characteristics (see panels (a-c) of Figure \ref{fig:9} where we label such a simulation as "no direct" and the parameters in Table S12). The explanation of the observed phenomena can be deduced from the set of various types of phonon DOS plots from Figures S57-S60 and those thermally-weighted presented in Figure \ref{fig:10}.
The ordinary DOS (Figure S57) and spin-phonon weighted (Figure S58) reveal that the states of the first excited doublet fall into the region of extremely low phonon density just below 40 cm$^{-1}$, above where many more optical vibrations appear. This is compensated for the low temperature of 5 K by phonon occupation as seen in Figure S59 and panel (a) of Figure \ref{fig:10} where the effective coupling is still stronger for the first excited doublet. This is exactly why we see the deviation of the model from the linear $\mathrm{ln}(\tau)$ vs $1/T$ dependency for the last point at 5 K excluded from the fitting (see panel (a) of Figure \ref{fig:9}). Then at higher temperature (Figure S60 and panel (b) of Figure \ref{fig:10}), the very efficient direct one-phonon excitation channel discussed around 70 cm$^{-1}$ dominates. Several factors may contribute to this discrepancy. Primarily, we can see that for an experimentally extracted energy barrier of 43.7(6) cm$^{-1}$, the coupling would be much stronger throughout the entire investigated temperature range as seen in Figure \ref{fig:10}, as we are very close to the steep cutoff around 40 cm$^{-1}$. We can therefore look for potential causes in our minimal CASSCF model containing only the first coordination sphere in the geometry of the host Y(III) - \textbf{4} lattice without treating the dynamical correlation that can give not precise enough energy splitting and coupling in tandem with pDFT calculations. In reality, the sole fact that phonon calculations and spin system modeling must be conducted using different levels of theory alone can lead to such energy mismatches. Another possible explanation is to postulate that the efficient direct one-phonon excitation channel is indeed quenched in real material due to the, e.g., hyperfine interactions with magnetically active $^{161}$Dy and $^{163}$Dy isotopes as theoretically investigated by some of us for Ho-based SMMs in \cite{Wang2021} and experimentally shown in \cite{Gonzalez2021}, where isotopic enrichment led to the first Yb-based zero-field SMM. As hyperfine interactions do not modify the energy splitting significantly, they can impact the quantum composition of states, often even leading to a change of their odd/even spin characters, which only strengthens the need for their future inclusion in our ab initio model. Moreover, we do not exclude here the potential impact of Raman-like relaxation, but looking at how well the expected static limits of susceptibility and temperature-dependent shapes of a.c. signals are recreated compared to Ab initio II and experiment (panels (b-c) of Figure \ref{fig:9}), those are not as probable as in \textbf{2}.

\begin{figure}[htbp!]
  \centering
  \includegraphics[scale=0.91]{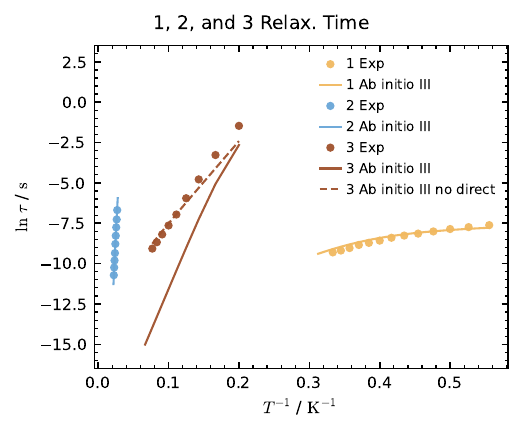}
  \caption{Temperature-variable experimental magnetic relaxation times (points) compared to the Ab initio III model results (lines).}
  \label{fig:11}
\end{figure}

Regardless, as shown in Figure \ref{fig:11},
even our minimal model, based on in silico modifications of host lattice \textbf{4} by substitution of various lanthanide ions, leading to \textbf{1}-\textbf{3} models, in combination with the developed linear response methodology, can correctly rank the effective temperature regions of a.c. response in a given frequency range between the investigated compounds. This further demonstrates the potential utility of the presented ab initio methods in the computational design of future SMMs, reducing unnecessary, tedious experimental labor in searching for many potential SMM systems, offering the possibility of getting theoretical insights into relaxation mechanisms, and ranking potential structural modifications completely in silico.

\section*{Summary and Conclusions}

In this work, we developed and implemented a first-principles linear-response methodology for open quantum systems that gives direct access to the complex a.c. magnetic susceptibility $\chi(\omega)$ of lanthanide-based molecular nanomagnets in the presence of an oscillating magnetic field. In contrast to the currently dominant master-equation strategies, which typically infer relaxation times from the eigenvalues of quantum master matrices. The present framework treats the experimentally measured quantity itself as the central object and establishes a direct bridge between microscopic spin-phonon coupling and recorded experimental $\chi'(\omega)$ and $\chi''(\omega)$. The formalism is based on the reduced response density operator and its exact Nakajima-Zwanzig equation of motion, specialized here in the linear vibronic coupling with a harmonic phonon bath, and cast into a Liouville-space matrix form suitable for numerical implementation. A major practical outcome is that the method retains explicit information about both the perturbing operator and the measured observable, and therefore about the relative visibility of different direction-dependent relaxation channels in powder a.c. measurements, which is lost in purely rate-based approaches.

Beyond the methodological advance itself, this contribution was designed as a self-contained theoretical reference. The main text, together with the extensive Supplementary Information, provides a unified introduction to a.c. susceptibility, phenomenological Debye/Havriliak-Negami analysis, closed- and open-system linear response, projection-operator techniques, harmonic lattice dynamics, linear spin-phonon coupling, and the ab initio evaluation of Hamiltonian gradients and spin-phonon operators, with detailed step-by-step derivations throughout. In this sense, this work also serves as a compact review and implementation-oriented guide for readers interested in understanding the full chain connecting microscopic Hamiltonians to simulated susceptibility curves.

On the computational side, we combined periodic phonon calculations for a three-dimensional coordination polymer host \textbf{4} with multiconfigurational relativistic SA-CASSCF calculations for embedded lanthanide clusters of \textbf{1}-\textbf{3}, obtaining spin-phonon coupling operators directly from gradients of the fully relativistic ab initio Hamiltonian without mapping onto crystal-field parameters or effective spin Hamiltonians. Together with the purpose-built Brillouin-zone multigrid, this enabled simulations of field- and temperature-dependent a.c. susceptibility in a genuinely periodic environment. For compound \textbf{1}, the method reproduces the direct-process regime and its field dependence in very good agreement with experiment, while for compound \textbf{2} it correctly captures the high-temperature one-phonon Orbach regime and clarifies the crossover to a lower-temperature Raman/LMP contribution. In both systems, the asymmetric broadening prescription implied by the $\omega \rightarrow 0$ limit of the response theory proved essential and consistently outperformed the conventional symmetric treatment. Compared with one-phonon Secular-Markov calculations, the present approach gives systematically improved agreement with experimental a.c. relaxation times and, importantly, yields the full powder-averaged susceptibility curves rather than only effective rates.

At the same time, the present study identifies the main challenges that remain to be solved. The complete treatment of the inhomogeneous term $\hat{M}_\mathrm{\psi}$ and the correlated initial condition $M_{\rho_A(0)}$ is numerically very demanding, especially if one wishes to later restore their full field frequency dependence. To partially address this, we demonstrated that we can effectively restore the expected susceptibility values employing a normalization procedure of results based on equations containing only the response memory kernel $\hat{M}_{\mathcal{K}_\mathrm{r}}$ and projected Liouvillian $M_\mathrm{L}$ matrices without altering the extracted relaxation times for investigated systems in any measurable way. However, the current second-order formulation is insufficient for systems and temperature ranges dominated by two-phonon Raman-type relaxation, as illustrated by compound \textbf{2}. The discrepancy found for compound \textbf{3} further indicates that quantitative predictions can still be limited by the quantum-chemical calculation accuracy caused by the finite geometry cluster approximation, the mismatch between electronic-structure and phonon levels of theory, and missing physical ingredients such as hyperfine interactions, dipolar fields, and additional dephasing channels. These points define a clear roadmap for future work: extension of the response formalism to fourth order in the system-bath coupling, inclusion of hyperfine- and dipolar-field effects relevant to QTM-like processes, and systematic refinement of the electronic and vibrational models. Nevertheless, the present results demonstrate that ab initio simulation of complex a.c. susceptibility in periodic lanthanide-based magnets is now feasible and can already provide both quantitative comparison with experiment and mechanistic insight into the occurring relaxation processes. We therefore expect the methodology developed here to become a useful tool for the fully in silico screening, ranking, and rational design of future single-molecule magnets.

\section*{Supplementary Information}
See the supplementary information for detailed derivations, expanded methods, and additional figures and tables.

\noindent
\textbf{Acknowledgements and Funding}\\
This research was financed by the National Science Center of Poland within the PRELUDIUM-20 project (grant no. 2021/41/N/ST4/04432). A.L. acknowledges funding from the European Research Council (ERC) under the European Union’s Horizon 2020 research and innovation programme (grant agreement No. [948493]). The study was partially carried out using the research infrastructure co-funded by the European Union in the framework of the Smart Growth Operational Program, Measure 4.2; Grant No. POIR.04.02.00-00-D001/20, “ATOMIN 2.0 – ATOMic scale science for the INnovative economy”. M.Z. and J.J.Z. acknowledge the support of the Foundation for Polish Science within a START fellowship 2025. The authors gratefully acknowledge Junhao Wang and Yue Xin for sharing unpublished data for compound \textbf{3} originating from our previous study. We further thank Lorenzo A. Mariano for helpful discussions and insightful suggestions concerning the implementation of the custom Hamiltonian-gradient methodology.

\bibliographystyle{achemso}
\bibliography{bibliography}

\end{document}